\documentclass[11pt]{article}
\usepackage{a4wide}
\usepackage{mathrsfs}
\usepackage{latexsym,bm}
\usepackage{graphicx}
\usepackage{indentfirst}
\usepackage{slashed}
\usepackage{amsmath}
\usepackage{amssymb}
\usepackage{xcolor}
\usepackage{hyperref}
\usepackage{epsfig}
\usepackage[titletoc]{appendix}
\usepackage{multirow}%
\usepackage{rotating}
\usepackage{cite}
\usepackage{ulem} 
\usepackage[top=2.9cm,bottom=2.5cm,left=2.8cm,right=3cm]{geometry}
\usepackage{tabularx}

\newcommand{\nn}{\nonumber}

\newcommand{\email}[1]{\footnote{{\em } \texttt{#1}}}

\newcommand{\mL}{\mathcal{L}}

\newcommand{\bra}{\langle}
\newcommand{\ket}{\rangle}
\newcommand{\xpt}{{\chi}{\rm PT}}

\begin{document}

\thispagestyle{empty}
\title{
\Large \bf Axion-meson mixing in light of recent lattice $\eta$-$\eta'$ simulations and their two-photon couplings within $U(3)$ chiral theory }
\author{\small Rui Gao$^a$,\, Zhi-Hui Guo$^{a,b}$\email{zhguo@hebtu.edu.cn}, \, J.~A.~Oller$^{c}$\email{oller@um.es},\, Hai-Qing Zhou$^{d}$\email{zhouhq@seu.edu.cn} \\[0.5em]
{ \small\it ${}^a$ Department of Physics and Hebei Key Laboratory of Photophysics Research and Application, } \\ 
{\small\it Hebei Normal University,  Shijiazhuang 050024, China}\\[0.2em] 
{ \small\it ${}^b$ CAS Key Laboratory of Theoretical Physics, Institute of Theoretical Physics, } \\
{\small \it Chinese Academy of Sciences, Beijing 100190, China}\\[0.2em]
{\small {\it ${}^c$ Departamento de F\'{\i}sica. Campus de Espinardo, Murcia E-30071, Spain}}\\[0.2em]
{ \small\it ${}^d$ School of Physics, Southeast University, Nanjing 211189, China }
}
\date{}

%

\maketitle
\begin{abstract}
We study the mixing of the QCD/QCD-like axion and light-flavor mesons $\pi^0, \eta, \eta'$ within the framework of $U(3)$ chiral perturbation theory up to next-to-leading order in this work. The axion-meson mixing formulas are calculated order by order in the $U(3)$ $\delta$-expansion scheme, namely the joint expansions of the momentum, light-quark masses and $1/N_C$. We provide  axion-meson mixing relations in terms of the $\pi^0$-$\eta$-$\eta'$ mixing parameters and their masses. The recent lattice simulations on the $\eta$-$\eta'$ systems turn out to be able to offer valuable inputs to constrain the unknown low-energy constants. The relation of the mass and decay constant of the axion is then further explored based on our updated calculations. The two-photon couplings of the light-flavor mesons, together with the axion, are also investigated in the $U(3)$ chiral theory up to next-to-leading order in the $\delta$-counting scheme.  
\end{abstract}

\section{Introduction}

The hypothetical particle axion provides an elegant solution to the long standing strong CP problem~\cite{Peccei:1977hh,Weinberg:1977ma,Wilczek:1977pj}. Since its first proposal in late 1970s, tremendous efforts have been put in various fields of physics to search this intriguing hypothesized  particle, ranging from particle physics, astronomy, cosmology to condensed matter and optical physics, etc~\cite{axionreviews}. 

In the Peccei-Quinn (PQ) picture, axion corresponds to the pseudo-Nambu-Goldstone boson (pNGB) resulting from the breaking of the global $U_{\rm PQ}(1)$ symmetry. In the low-energy regime, the most characteristic interaction of the axion is the coupling with the topological gluon density, i.e., $a/f_a\,G^{\mu\nu}\tilde{G}_{\mu\nu}$, where $G^{\mu\nu}$ is the gluon field strength tensor, its dual is $\tilde{G}_{\mu\nu}=\varepsilon_{\mu\nu\rho\sigma}G^{\rho\sigma}/2$  with $\varepsilon_{\mu\nu\rho\sigma}$ the Levi-Civita antisymmetric tensor and $f_a$ stands for the axion decay constant. The essential point of the PQ mechanism to solve the strong CP problem relies on the cancellation of the CP violating term $\theta G^{\mu\nu}\tilde{G}_{\mu\nu}$ in QCD by the dynamical generation of a proper vacuum expectation value (VEV) from the axion field. Different ultraviolet (UV) models in the new physics sector can lead to rather different couplings of the axion with other standard-model (SM) particles, such as the leptons, quarks, photon and W/Z bosons~\cite{Weinberg:1977ma,Wilczek:1977pj,Dine:1981rt,Zhitnitsky:1980tq,Kim:1979if,Shifman:1979if}, although they can be constructed to give a universal $G\tilde{G}$ coupling. 

Due to the nonperturbative nature of the gluons in the low-energy region, the common way to study the interactions of the axion is to first perform the axial transformation of the quark fields to eliminate the $a/f_a\,G^{\mu\nu}\tilde{G}_{\mu\nu}$ operator from the beginning and then match to the axion chiral perturbation theory ($\chi$PT), which encodes the axion field together with the pions in the $SU(2)$ case, and the octet of $\pi, K, \eta_8$ in the $SU(3)$ case. On the other hand, it is also possible to directly match the anomalous axion term $a/f_a\,G^{\mu\nu}\tilde{G}_{\mu\nu}$ with the $\xpt$ operators. In this regard, it is reminded that the large mass of the singlet pseudoscalar $\eta_0$ (the dominant component of the physical $\eta'$ meson), compared to those of the pseudoscalar octet $\pi, K$ and $ \eta_8$, can be mainly attributed to the anomalous breaking of the QCD $U_A(1)$ symmetry. Therefore, it implies that another way to introduce the axion field into $\chi$PT can be similar to the case of the singlet $\eta_0$. The pioneer work to study the influence of the $\eta'$ on the axion properties can trace back to Refs.~\cite{Choi:1986zw,Spalinski:1988yf}. Recently many works aiming at the extended descriptions of the axion-$\eta^{(')}$ interactions are proposed from different points of view~\cite{Alves:2020xhf,Bigazzi:2019hav,DiVecchia:2017xpu,Ertas:2020xcc,Gan:2020aco,Landini:2019eck,Bottaro:2020dqh,REDTOP:2022slw,Aloni:2018vki,Alves:2017avw}, including the axion production from the $\eta/\eta'$ and kaon decays, CP violating axion signals, the axion-baryon couplings, the model-independent part of the axion-photon-photon coupling, etc.  
In this work, we will further pursue a systematical calculation of the axion-$\phi$ ($\phi=\pi, \eta, \eta'$) interactions order by order within the $U(3)$ $\xpt$ by employing the $\delta$ expansion scheme~\cite{Herrera-Siklody:1996tqr,Leutwyler:1997yr,Kaiser:2000gs}, namely, a simultaneous expansion in powers of the momenta, light-quark masses and $1/N_C$, i.e. $\delta \sim p^2 \sim m_q \sim 1/N_C$. In addition, we also study the relations of the model-independent two-photon couplings of the axion, together with those of the $\pi^0,\eta$ and $\eta'$ mesons, by performing the next-to-leading order (NLO) $U(3)$ $\xpt$ calculations. 

The layout of this paper is as follows. The relevant chiral Lagrangians up to NLO are elaborated in Sec.~\ref{sec.lag}. We address the mixing formalism at leading order (LO) for the four-particle system involving $\pi^0, \eta, \eta'$ and the axion in Sec.~\ref{sec.mixlo}. The calculation of the NLO mixing, the fits to relevant lattice data and numerical analyses of the transitions matrix elements, and the masses of $\pi^0, \eta, \eta'$ and the axion are given in Sec.~\ref{sec.mixnlo}. The two-photon couplings of the aforementioned four particles are then studied up to NLO in the $U(3)$ $\chi$PT in Sec.~\ref{sec.twophoton}. A short summary and conclusions are given in Sec.~\ref{sec.summary}.

\section{Axion $U(3)$ $\xpt$ up to next-to-leading order}\label{sec.lag}

The axion is characterized by the interaction with the gluons and its effective operators can be written as 
\begin{eqnarray}\label{eq.lagag}
\mathcal{L}_a^{G}= \frac{1}{2}\partial_\mu a \partial^\mu a + \frac{a}{f_a} \frac{\alpha_s}{8\pi} \sum_{i=1}^{8} G^i_{\mu\nu}\tilde{G}^{i,\mu\nu}  - \frac{1}{2} m_{a,0}^{2}  \, a^2 \,,
\end{eqnarray}
where $G^i_{\mu\nu}$ and $\tilde{G}^i_{\mu\nu}$ are the gluon field strength tensor and its dual, with $i$ the color indices. The second term in Eq.~\eqref{eq.lagag} is considered to be model-independent, due to its relevance of solving the strong CP problem. For the bare mass $m_{a,0}$ of the axion in the third term, its value is usually assumed to be vanishing in the minimal QCD axion  setup~\cite{Peccei:1977hh,Weinberg:1977ma,Wilczek:1977pj}, although it is also possible to have a nonvanishing $m_{a,0}$ to solve the strong CP problem~\cite{Dimopoulos:1979pp,Holdom:1982ex,Rubakov:1997vp,Gaillard:2018xgk,axionreviews}. The direct couplings of the axion with the photon and fermions heavily rely on the specific model-building considerations in the BSM sector. In this work, we focus on the axion interactions with the hadrons and photon that are purely induced by the effective Lagrangian in Eq.~\eqref{eq.lagag}, i.e. the model-independent parts of the axion-hadron and axion-photon interactions. 

The next step is to match the effective Lagrangian in Eq.~\eqref{eq.lagag} to the axion chiral perturbation theory~\cite{Georgi:1986df,GrillidiCortona:2015jxo}, which provides a reliable framework to study the axion-hadron interactions order by order. In the low-energy QCD, apart from the chiral symmetry breaking, another distinct feature is the QCD $U_A(1)$ anomaly, i.e. the anomalous breaking of the $U_A(1)$ symmetry by topological charge density $\omega(x)=\alpha_s G_{\mu\nu}\tilde{G}^{\mu\nu}/(8\pi)$, which gives a natural explanation of the large mass of the singlet $\eta_0$ even in the chiral limit. 
In this work we will carry out the study sticking to the $U(3)$ $\xpt$ by employing the large $N_C$ argument~\cite{ua1nc}. This implies that one can explicitly include the axion field into the $\xpt$ Lagrangian in a similar way as the situation of the $\eta'$. 
To be more specific, we use the $\delta$-expansion scheme to arrange the various contributions in $U(3)$ $\xpt$ by simultaneously considering the expansions in the momenta, light-quark masses and $1/N_C$~\cite{Leutwyler:1997yr,Kaiser:2000gs,Herrera-Siklody:1996tqr}. 

To set up the notations, we briefly recapitulate the way to construct the LO operators of $U(3)$ $\xpt$. From the large $N_C$ point of view, the singlet $\eta_0$ would become the ninth pNGB in the large $N_C$ and chiral limits, since the instanton effect via the QCD $U_A(1)$ anomaly is $1/N_C$ suppressed. Then, the dynamical fields in low energy QCD form the pNGB nonet, which is usually parameterized as $U={\exp}\big(i\sqrt2\Phi/F\big)$, with the pNGB nonet matrix given by
\begin{equation}\label{phi1}
\Phi \,=\, \left( \begin{array}{ccc}
\frac{1}{\sqrt{2}} \pi^0+\frac{1}{\sqrt{6}}\eta_8+\frac{1}{\sqrt{3}} \eta_0 & \pi^+ & K^+ \\ \pi^- &
\frac{-1}{\sqrt{2}} \pi^0+\frac{1}{\sqrt{6}}\eta_8+\frac{1}{\sqrt{3}} \eta_0   & K^0 \\  K^- & \overline{K}^0 &
\frac{-2}{\sqrt{6}}\eta_8+\frac{1}{\sqrt{3}} \eta_0
\end{array} \right)\,.
\end{equation}
The QCD $U_A(1)$ anomaly effect can be incorporated in effective Lagrangian via the operator $-\frac{\tau}{2} ( -i \log\det U )^2$,  
where $\tau$ corresponds to the topological susceptibility~\cite{Kaiser:2000gs}. Taking into account that $\det U=\exp({\rm Tr}\log U)$, the former operator can be cast as $-3\tau \eta_0^2/{F^2}$, which is nothing but the mass term for singlet $\eta_0$ in the chiral limit. In practice it is convenient to introduce $M_0^2=\frac{6\tau}{F^2}$ as the LO mass squared for $\eta_0$. In the large $N_C$ counting rule, the topological susceptibility $\tau$ and the pNGB decay constant $F$ scale like $O(1)$ and $O(\sqrt{N_C})$, respectively~\cite{Kaiser:2000gs,Gasser:1984gg}, which indicate that $M_0^2$ behaves as $O(1/N_C)$. Under the PQ assumption, the CP violating $\theta G\tilde{G}$ term is canceled by the VEV part of the axion via the $aG\tilde{G}$ term in Eq.~\eqref{eq.lagag}. Therefore, this indicates that the anomalous $aG\tilde{G}$ effect can be included in the chiral effective Lagrangian in a similar fashion as the $\theta$ term, the latter of which is discussed in great detail in Ref.~\cite{Kaiser:2000gs}. According to the recipe in the previous reference, the axion field can be introduced to the LO $U(3)$ $\xpt$ together with the anomalous mass for $\eta_0$ via
$-\frac{\tau}{2} ( -i \log\det U + a/f_a )^2=-\frac{\tau}{2} ( \sqrt{6}\eta_0/F + a/f_a )^2$. For phenomenological convenience, we will always use the parameter $M_0$ to replace the topological susceptibility $\tau$ in the following discussions. On general grounds, the axion chiral transformation on the quark fields needed to remove the $aG\tilde{G}$ term in Eq.~\eqref{eq.lagag} is of the same type as the singlet axial chiral transformation $U_A(1)$ that is  parameterized as $\exp(i\eta_0\sqrt{2/3})$. This observation drives to the necessity of adding together the fields of the axion and $\eta_0$ in the large $N_C$ and chiral limit, where the latter is a pNGB of QCD parameterized as coordinates in the  coset space $U_L(3)\otimes U_R(3)/U_V(3)$ \cite{coleman69}.  Under these circumstances, the LO $U(3)$ $\xpt$ including the axion field takes the form 
\begin{eqnarray}\label{eq.laglo}
\mL^{\rm LO}= \frac{ F^2}{4}\bra u_\mu u^\mu \ket+
\frac{F^2}{4}\bra \chi_+ \ket
+ \frac{F^2}{12}M_0^2 X^2 \,,
\end{eqnarray}
where the axion field is introduced via 
\begin{eqnarray}\label{eq.defx}
X= \log{(\det U)} + i\frac{a}{f_a}\,,  
\end{eqnarray}
and other basic $\xpt$ building blocks are given by 
\begin{eqnarray}\label{defbb}
&& U =  u^2 = e^{i\frac{ \sqrt2\Phi}{ F}}\,, \qquad \chi = 2 B (s + i p) \,,\qquad \chi_\pm  = u^\dagger  \chi u^\dagger  \pm  u \chi^\dagger  u \,,
  \nn\\
&& u_\mu = i u^\dagger  D_\mu U u^\dagger \,,  \qquad   D_\mu U \, =\, \partial_\mu U - i (v_\mu + a_\mu) U\, + i U  (v_\mu - a_\mu) \,.
\end{eqnarray}
Here, $s$, $p$, $v_\mu$, and $a_\mu$ are external scalar, pseudoscalar, vector and axial-vector external sources, respectively, introduced as spurion fields. 
The quark-mass corrections are introduced by fixing $s=M_q\equiv{\rm diag}(m_u,m_d,m_s)$. The parameter $M_0$ in the last term of Eq.~\eqref{eq.laglo} corresponds to the leading-order mass of the $\eta_0$, which squared is proportional to the topological susceptibility~\cite{Kaiser:2000gs}. The leading scaling behavior of $M_0^2$ is $O(1/N_C)$ in the $\delta$-counting scheme~\cite{ua1nc}. 

Another approach frequently used in literature to introduce the axion field in $\xpt$, e.g. as those in Refs.~\cite{Georgi:1986df,GrillidiCortona:2015jxo},  is to first perform the axial transformation of the quark fields to eliminate the axion-gluon operator in Eq.~\eqref{eq.lagag}, i.e., by taking 
\begin{eqnarray}\label{eq.qtransf}
q \to e^{i\frac{a}{2f_a} \gamma_5  Q_a} q\,, \qquad [ q=(u,d,s)^{\rm T} ]\,,
\end{eqnarray}
where $Q_a$ is a $3\times 3$ matrix spanned in the light three-flavor space. Such a transformation induces two additional terms in the effective Lagrangian~\eqref{eq.lagag}: $-\frac{a \alpha_s}{8\pi f_a} G_{\mu\nu} \tilde{G}^{\mu\nu} \, {\rm Tr}(Q_a) $ and $-\frac{\partial_\mu a}{2f_a} \bar{q} \gamma^\mu \gamma_5 Q_a q$. The former term exactly cancels the axion-gluon operator in Eq.~\eqref{eq.lagag}, after taking the constraint ${\rm Tr}(Q_a)=1$. While, the latter term describes the extra axion-quark interaction that is induced by the axial transformation~\eqref{eq.qtransf}. In addition, it is easy to demonstrate that this transformation will also introduce the axion field into the quark masses: $M_q \to M_q(a)= e^{i \frac{a}{2f_a} Q_a} M_q e^{i \frac{a}{2f_a}Q_a}$, which can lead to non-derivative axion interactions. In the $SU(3)$ $\chi$PT, the LO mass mixing between the axion and the neutral unflavored $\pi^0$ and $\eta_8$ can be avoided by taking $Q_a\propto M_q^{-1}$. However, in the $U(3)$ case, even if one imposes the latter form for $Q_a$, the mixing between the axion and the singlet $\eta_0$ still exists at LO. Therefore, in this work we will not perform the transformation of quark fields~\eqref{eq.qtransf}, and use instead the original $U(3)$ Lagrangian in Eq.~\eqref{eq.laglo} to proceed the calculations. The physical results should be the same regardless of keeping the $aG\tilde{G}$ term or replacing it via the axial transformation~\eqref{eq.qtransf}, although the intermediate steps in the $\xpt$ calculations can look different.

In the rest of the discussions, we will stick to the method by including the axion in the $U(3)$ $\xpt$ through the $X$ field of Eq.~\eqref{eq.defx}. In this case, the axion $U(3)$ $\xpt$ Lagrangian coincides with the standard one with the obvious replacement of the $X$ field.  When restricting to the axion-meson mixing, the relevant NLO Lagrangian under the $\delta$-counting rule consists of four operators 
\begin{eqnarray}\label{eq.lagnlo}
\mL^{\rm NLO} =  L_5 \bra  u^\mu u_\mu \chi_+ \ket
+\frac{ L_8}{2} \bra  \chi_+\chi_+ + \chi_-\chi_- \ket
-\frac{F^2\, \Lambda_1}{12}   D^\mu X D_\mu X  -\frac{F^2\, \Lambda_2}{12} X \bra \chi_- \ket\,, 
\end{eqnarray}
where the first two terms accompanied by $L_5$ and $L_8$ share the same forms as those from the conventional $SU(3)$ $\xpt$~\cite{Gasser:1984gg}. Within the framework of $\xpt$, the $N_C$ order of a given operator can be inferred by the number of traces in the flavor space~\cite{Gasser:1984gg}. Generally speaking, one extra trace will bring one additional $1/N_C$ suppression order to the effective operator. By taking into account the identity of $\log\det U= \bra \log U \ket$, it is straightforward to conclude that the first two terms in Eq.~\eqref{eq.lagnlo} are counted as $O(p^4,N_C)$, and the remaining two terms with $\Lambda_1$ and $\Lambda_2$ that only appear in the $U(3)$ case~\cite{Kaiser:2000gs} are counted as $O(p^2,N_C^0)$.

The two-photon decays of the light pseudoscalar mesons are governed by the Wess-Zumino-Witten Lagrangian and the relevant LO Lagrangian~\cite{Wess:1971yu,Witten:1983tw} is 
\begin{eqnarray}\label{eq.lagwzwlo}
\mathcal{L}_{WZW}^{\rm LO}= -\frac{3\sqrt{2}e^2}{8\pi^2 F}\varepsilon_{\mu\nu\rho\sigma}\partial^\mu A^\nu \partial^\rho A^\sigma \bra Q^2 \Phi \ket  \,,
\end{eqnarray}
which is counted as $O(p^4,N_C)$ in the $\delta$-counting scheme. The quantity $Q$ stands for the matrix of the electric charges of the light quarks, i.e. $Q={\rm Diag}(\frac{2e}{3},-\frac{e}{3},-\frac{e}{3})$, with $e$ the magnitude of the electron charge, and $A_\mu$ stands for the photon field. At NLO there are two additional terms~\cite{Moussallam:1994xp,Bijnens:2001bb} 
\begin{eqnarray} \label{eq.lagwzwnlo}
\mathcal{L}_{WZW}^{\rm NLO}=  i t_1 \varepsilon_{\mu\nu\rho\sigma}  \bra  f_+^{\mu\nu} f_+^{\rho\sigma} \chi_- \ket  +  k_3 \varepsilon_{\mu\nu\rho\sigma} \bra  f_+^{\mu\nu} f_+^{\rho\sigma}  \ket  X \,,  
\end{eqnarray}
with 
\begin{eqnarray}
 f_+^{\mu\nu}= u F_L^{\mu\nu} u^\dagger + u^\dagger F_R^{\mu\nu} u\,,
\end{eqnarray}
where $F_L^{\mu\nu}=\partial^\mu F_L^\nu-\partial^\mu F_L^\nu-i[F_L^\mu,F_L^\nu]$ and $F_R^{\mu\nu}=\partial^\mu F_R^\nu-\partial^\mu F_R^\nu-i[F_R^\mu,F_R^\nu]$, with $F_L^\mu=v^\mu-a^\mu$ and $F_R^\mu=v^\mu+a^\mu$, denote the field strength tensors of left-hand and right-hand external sources, respectively. One can take $F_L^\mu$=$F_R^\mu$= $-e Q A^\mu$ to obtain the interaction vertexes with photons. 
The $t_1$ and $k_3$ terms in Eq.~\eqref{eq.lagwzwnlo} belong to the $O(p^6,N_C)$ and $O(p^4,N_C^0)$ types of operators in the $\delta$ counting, respectively. When restricting to the pseudoscalar-photon-photon case, the Lagrangian~\eqref{eq.lagwzwnlo} reduces to 
\begin{align}\label{eq.lagwzwnloexpd}
\mathcal{L}_{WZW}^{\rm NLO}=&  t_1 \frac{32\sqrt{2} B }{F}\varepsilon_{\mu\nu\rho\sigma}\partial^\mu A^\nu \partial^\rho A^\sigma \bra \big( M_q \Phi + \Phi M_q  \big) Q^2 \ket \nonumber \\ &+
   16  k_3 \varepsilon_{\mu\nu\rho\sigma}\partial^\mu A^\nu \partial^\rho A^\sigma \bra Q^2 \ket \bigg( \frac{\sqrt{2}}{F} \bra \Phi \ket + \frac{a}{f_a} \bigg)\,.
\end{align}
According to the Lagrangians in Eqs.~\eqref{eq.laglo} and \eqref{eq.lagwzwlo}, the LO interactions of the axion with the pNGBs and photons are purely caused by the mixing of the axion and the neutral unflavored pNGBs. Therefore it is mandatory to first address the axion-pNGBs mixing problem.

\section{Revisiting the mixing formalism at LO}\label{sec.mixlo}

In principle, there are both kinetic and mass mixings between different states \cite{Jamin:2000wn,Jamin:2001zq}. 
At LO in the $\delta$ counting, the $\pi^0, \eta, \eta'$ and $a$ will get mixing purely via the mass terms from the Lagrangian~\eqref{eq.laglo}, while the kinetic mixing only starts to appear in the NLO Lagrangian~\eqref{eq.lagnlo}. We calculate the mixing of the axion and $\pi^0, \eta, \eta'$ order by order below. 

In Refs.~\cite{Guo:2011pa,Guo:2012ym,Guo:2012yt,Guo:2015xva}, the $\eta$-$\eta'$ mixing has been studied up to next-to-next-to leading order in the $\delta$ expansion in the isospin limit. As first pointed out in Ref.~\cite{Guo:2011pa}, it is convenient to use the $\mathring{\overline{\eta}}$ and $\mathring{\overline{\eta}}'$ that are diagonalized at LO, if one attempts to perform a systematical higher-order calculation in the $U(3)$ chiral theory relying on the $\delta$ counting. The key reason is that the LO mixing vertex of the $\eta_8$ and $\eta_0$ is formally counted as the same order of their LO masses, implying that an infinity insertion of the LO mixing vertex of $\eta_0$ and $\eta_8$ in the loop calculations will not get suppressed in the $\delta$-counting scheme, see Fig.~1 of Ref.~\cite{Guo:2011pa} for illustrations. This cumbersome problem can be avoided by first performing the LO diagonalization of $\eta_8$ and $\eta_0$ and using the LO diagonalized fields $\mathring{\overline{\eta}}$ and $\mathring{\overline{\eta}}'$, whose relations are given by
\begin{eqnarray}\label{eq.lomixing}
\left(\begin{array}{c} \mathring{\overline{\eta}} \\ \mathring{\overline{\eta}}' \end{array}\right) =
\left(\begin{array}{cc} c_\theta & - s_\theta \\  s_\theta & c_\theta \end{array}\right)  
\left(\begin{array}{c} \eta_8 \\ \eta_0\end{array}\right) \, ,
\end{eqnarray}
with $c_\theta=\cos{\theta}$ and $s_\theta=\sin{\theta}$. 
The reason behind is that the mixing strengths between $\mathring{\overline{\eta}}$ and $\mathring{\overline{\eta}}'$ are suppressed at least by one higher order in the $\delta$-counting rule.  
From the Lagrangian \eqref{eq.laglo}, the LO mixing angle $\theta$
and masses of $\mathring{\overline{\eta}}$ and $\mathring{\overline{\eta}}'$ can be calculated  
\begin{eqnarray} 
m_{\overline{\eta}}^2 &=& \frac{M_0^2}{2} + m_{\overline{K}}^2
- \frac{\sqrt{M_0^4 - \frac{4 M_0^2 \Delta^2}{3}+ 4 \Delta^4 }}{2} \,, \label{eq.defmetab2}  \\
m_{\overline{\eta}'}^2 &=& \frac{M_0^2}{2} + m_{\overline{K}}^2
+ \frac{\sqrt{M_0^4 - \frac{4 M_0^2 \Delta^2}{3}+ 4 \Delta^4 }}{2} \,, \label{eq.defmetaPb2}  \\
\sin{\theta} &=& -\left( \sqrt{1 +
\frac{ \big(3M_0^2 - 2\Delta^2 +\sqrt{9M_0^4-12 M_0^2 \Delta^2 +36 \Delta^4 } \big)^2}{32 \Delta^4} } ~\right )^{-1}\,,
\label{eq.deftheta0}
\end{eqnarray}
where $\Delta^2 = m_{\overline{K}}^2 - m_{\overline{\pi}}^2$, and $m_{\overline{\pi}}$ and $m_{\overline{K}}$ are the LO masses of the pion and kaon in the isospin symmetric limit, respectively.

New subtle complexities will arise when simultaneously including the $\pi^0$, the axion $a$, $\eta$ and $\eta'$. On  one hand, the isospin breaking (IB) contributions can not be ignored any more, since the mixings  between the $\pi^0$ and the remaining fields $\eta$, $\eta'$ and $a$ are caused by the IB effects. It is noted that we only consider the IB effects arising from the strong interaction, namely the mass difference between the up and down quarks. On the other hand, the mixings between the axion $a$ and other fields $\pi^0$, $\eta$ and $\eta'$ are suppressed by the factor $F/f_a$. Additional power counting rules will be introduced to simplify the discussions, apart from the conventional $\delta$-counting that is widely employed in the $\eta$-$\eta'$ case. Generally speaking, the magnitudes of the IB corrections are around $1\sim 2 \%$. Compared to the correction in the $\delta$ counting, which can be naively estimated to be around $30\%$ [$1/(N_C=3)$], it is justified to just keep the leading nonvanishing IB terms. The magnitude of the axion-related suppression factor $F/f_a$ is definitely much more smaller than that of the IB corrections. Therefore, at the current stage, it is safe to just take the leading $F/f_a$ terms. More precisely, we aim at the systematical calculations of the higher-order corrections in the $\delta$-counting scheme by keeping the leading IB and $F/f_a$ terms in this work. 

Similarly as the $\eta$-$\eta'$ case, it is much more convenient to use the fields $\overline{\pi}^0$, $\overline{\eta}$, $\overline{\eta}'$ and $\overline{a}$ that already diagonalize LO mass term to systematically calculate the higher-order corrections. Notice that we have used $\mathring{\overline{\eta}}$ and $\mathring{\overline{\eta}}'$ to denote the LO diagonalized $\eta$ and $\eta'$ fields in the isospin symmetric case, while $\overline{\eta}$ and $\overline{\eta}'$ stand for the fields after including the IB effects. As stated before, the leading terms in the IB and $F/f_a$ expansions will be kept in our calculation and, in this case, one can obtain the diagonalized states $\overline{\pi}^0$, $\overline{\eta}$, $\overline{\eta}'$ and $\overline{a}$ by performing the following field redefinitions
\begin{equation}\label{eq.mixlo08} 
\left( \begin{array}{c}
\overline{\pi}^0 \\  \overline{\eta} \\ \overline{\eta}' \\ \overline{a} 
\end{array} \right) \,=\, \left( \begin{array}{cccc}
1+O(v^2) & -v_{12} & -v_{13} & -v_{14} \\ 
v_{12} & 1+O(v^2) & -v_{23} & -v_{24}  \\ 
v_{13} & v_{23} & 1+O(v^2) & -v_{34} \\ 
v_{14} & v_{24} & v_{34} & 1+O(v^2) 
\end{array} \right)\,
\left( \begin{array}{c}
\pi^0 \\  \mathring{\overline{\eta}} \\ \mathring{\overline{\eta}}' \\ a
\end{array} \right) \,,
\end{equation}
where $\pi^0$ and $a$ correspond to the bare states entering in the  Lagrangian~\eqref{eq.laglo}. The matrix elements $v_{ij}$ are determined by the mass mixing terms from the LO Lagrangian~\eqref{eq.laglo}, and their explicit expressions are found to be
\begin{eqnarray}\label{eq.v12}
v_{12}=& -\frac{\epsilon}{\sqrt{3}} \frac{c_\theta-\sqrt{2}s_\theta}{m_{\overline{\pi}}^2-m_{\overline{\eta}}^2}\,, \\
v_{13}=& -\frac{\epsilon}{\sqrt{3}} \frac{\sqrt{2}c_\theta+s_\theta}{m_{\overline{\pi}}^2-m_{\overline{\eta}'}^2}\,, \\
v_{14}=& \frac{M_0^2\epsilon}{3\sqrt{2}(m_{a,0}^2-m_{\overline{\pi}}^2)} \frac{F}{f_a}\bigg[ \frac{s_\theta(c_\theta-\sqrt{2}s_\theta)(m_{a,0}^2+m_{\overline{\pi}}^2-m_{\overline{\eta}}^2)}{(m_{a,0}^2-m_{\overline{\eta}}^2)(m_{\overline{\eta}}^2-m_{\overline{\pi}}^2)}-\frac{c_\theta(\sqrt{2}c_\theta+s_\theta)(m_{a,0}^2+m_{\overline{\pi}}^2-m_{\overline{\eta}'}^2)}{(m_{a,0}^2-m_{\overline{\eta}'}^2)(m_{\overline{\eta}'}^2-m_{\overline{\pi}}^2)} \bigg]\,, \nonumber \\ \label{eq.v14}\\ \label{eq.v24}
v_{24}=& -\frac{M_0^2 s_\theta}{\sqrt{6}(m_{a,0}^2-m_{\overline{\eta}}^2)} \frac{F}{f_a}\,,\\ \label{eq.v34}
v_{34}=& \frac{M_0^2 c_\theta}{\sqrt{6}(m_{a,0}^2-m_{\overline{\eta}'}^2)} \frac{F}{f_a}\,,
\end{eqnarray}
where $m_{\overline{\pi}}$, $m_{\overline{\eta}}$ and $m_{\overline{\eta}'}$ correspond to the LO masses of the $\pi$, $\eta$ and $\eta'$ mesons, and $\epsilon=B(m_u - m_d)$ corresponds to the leading strong IB factor. Regarding the matrix element $v_{23}$ in Eq.~\eqref{eq.mixlo08}, it describes the IB contribution to the $\eta$-$\eta'$ mixing, which is expected to play negligible roles comparing with the $SU(3)$ symmetry breaking effects. Therefore in this work we will neglect the IB contributions to the $\eta$-$\eta'$ mixing, i.e., $v_{23}$ will be set to zero throughout.  
At the LO electromagnetic (EM) correction, the Dashen's theorem tells us that the EM corrections to the kaon masses equal to the mass differences of the pions~\cite{Dashen:1969eg,Bijnens:2001bb}, i.e. $(m_{K^{+}}^2 - m_{K^0}^2)_{\rm EM} = m_{\pi^{+}}^2 - m_{\pi^0}^2$, and at this order one can write  $\epsilon \equiv B(m_u - m_d) = m_{K^{+}}^2 - m_{K^0}^2 - (m_{\pi^{+}}^2 - m_{\pi^0}^2)$.

After the diagonalization at LO, the mass of the axion field $\overline{a}$ is found to be 
\begin{eqnarray}\label{eq.massaxion}
m_{\overline{a}}^2 = m_{a,0}^2 + \frac{M_0^2 F^2}{6 f_a^2} \bigg[ 1 + \frac{c_\theta^2 M_0^2 (2m_{a,0}^2 - m_{\overline{\eta}'}^2)}{(m_{a,0}^2 - m_{\overline{\eta}'}^2)^2} +  \frac{s_\theta^2 M_0^2 (2m_{a,0}^2 - m_{\overline{\eta}}^2)}{(m_{a,0}^2 - m_{\overline{\eta}}^2)^2} \bigg] + O(\epsilon)\,,
\end{eqnarray}
which  by taking $m_{a,0}=0$ reduces to the minimal setup of the QCD axion case 
\begin{eqnarray}\label{eq.massaxionqcd}
m_{\overline{a}}^2 =  \frac{M_0^2 F^2}{6 f_a^2} \bigg[ 1-  \frac{c_\theta^2 M_0^2 }{m_{\overline{\eta}'}^2} - \frac{s_\theta^2 M_0^2  }{ m_{\overline{\eta}}^2 } \bigg] + O(\epsilon)\,.
\end{eqnarray}
It should be stressed that the neat analytical expressions in Eqs.~\eqref{eq.v12}-\eqref{eq.massaxionqcd} are obtained by keeping the leading IB effects in $\pi^0$-$a$ and $\pi^0-\eta^{(')}$ mixing and neglecting the IB contributions to the $a$-$\eta^{(')}$ mixing and the masses of the axion and pNGBs.  
The concise formulas in Eqs.~\eqref{eq.massaxion} and \eqref{eq.massaxionqcd} give us direct access to estimate the influences from the pertinent $\eta$-$\eta'$ mixing parameters to the axion mass in different scenarios. We take the minimal QCD axion case, i.e. Eq.~\eqref{eq.massaxionqcd}, to carry out some exploratory phenomenological discussions.  

By taking the LO expressions of the $\eta$-$\eta'$ mixing in  Eqs.~\eqref{eq.defmetab2}, \eqref{eq.defmetaPb2} and \eqref{eq.deftheta0}, it is easy to demonstrate that the axion mass squared can be recast as 
\begin{eqnarray}\label{eq.massaxionqcd1}
m_{\overline{a}}^2 = \frac{F^2}{f_a^2} \frac{M_0^2 m_{\overline{\pi}}^2(2m_{\overline{K}}^2-m_{\overline{\pi}}^2)}{8M_0^2m_{\overline{K}}^2-2m_{\overline{\pi}}^2(M_0^2-6m_{\overline{K}}^2)-6m_{\overline{\pi}}^4}\,,
\end{eqnarray} 
which reduces to the well celebrated result $m_{\overline{a}}^2= F^2 m_{\overline{\pi}}^2/(4f_a^2)$~\cite{Weinberg:1977ma} by keeping the leading expansions of $m_{\overline{\pi}}^2/m_{\overline{K}}^2$ and $m_{\overline{\pi}}^2/M_0^2$.

For the LO fit in Ref.~\cite{Gu:2018swy}, the value of $M_0$ is determined to be $M_0=820.0$~MeV, which leads to 
\begin{eqnarray}\label{eq.lomassetetp}
m_{\overline{\eta}}=493.6~{\rm MeV},\quad m_{\overline{\eta}'}=957.7~{\rm MeV}, \quad \theta=-19.6^{\circ}\,. 
\end{eqnarray}
By substituting these values to Eqs.~\eqref{eq.v14}-\eqref{eq.massaxion} with $m_{a,0}=0$, we obtain  
\begin{eqnarray}\label{eq.malo1}
 m_{\overline{a}}= 6.1~{\rm \mu eV} \frac{10^{12}{\rm GeV}}{f_a}~\,, \quad v_{14}= \frac{-0.012}{f_a}\,,\quad v_{24}= \frac{-0.035}{f_a}\,,\quad v_{34}= \frac{-0.026}{f_a}\,,
\end{eqnarray}
where the mass agrees well with the recent determinations~\cite{GrillidiCortona:2015jxo,Lu:2020rhp}. The units of the numbers in the numerators of $v_{14/24/34}$ are taken as GeV here. Nevertheless, it is clear that the LO predictions for the masses of $\eta$ and $\eta'$ in Eq.~\eqref{eq.lomassetetp} are still not satisfactory when compared to their physical values. In naively estimating $m_{\overline{\eta}}$ and $m_{\overline{\eta}'}$ by the  physical masses of $\eta$ and $\eta'$, the axion mass can be predicted from Eq.~\eqref{eq.massaxionqcd} with the result 
\begin{eqnarray}\label{eq.malo2}
  m_{\overline{a}}= 9.7~{\rm \mu eV} \frac{10^{12}{\rm GeV}}{f_a}~\,, \quad v_{14}= \frac{-0.011}{f_a}\,,\quad v_{24}= \frac{-0.028}{f_a}\,,\quad v_{34}= \frac{-0.026}{f_a}\,,
\end{eqnarray}
with $M_0=820$~MeV and $\theta=-19.6^\circ$. In turn, if we take $M_0=820$~MeV and $\theta=-10.0^\circ$ the result turns out to be 
\begin{eqnarray}\label{eq.malo3}
  m_{\overline{a}}= 14.5~{\rm \mu eV} \frac{10^{12}{\rm GeV}}{f_a}~\,,  \quad v_{14}= \frac{-0.008}{f_a}\,,\quad v_{24}= \frac{-0.015}{f_a}\,,\quad v_{34}= \frac{-0.027}{f_a}\,.
\end{eqnarray}
The obvious changes of the predictions in Eqs.~\eqref{eq.malo1}-\eqref{eq.malo3} by taking different phenomenological inputs for the $m_{\bar{\eta}}$ and $m_{\bar{\eta}'}$, give hints that there could be potentially noticeable higher order corrections. This urges us to further continue the NLO calculations and verify their impacts on the axion properties.

\section{Mixing at NLO and Phenomenological discussions}\label{sec.mixnlo}

Up to the NLO in the $\delta$ counting, the bilinear terms involving the $\overline{\pi}^0,\overline{\eta},\overline{\eta}'$ and the axion field $\overline{a}$, which are diagonalized at LO, can be generally written as 
\begin{equation}\label{eq.lagsenlo}
\begin{aligned}
\mathcal{L}=&\dfrac{1+\delta^{\eta}_k}{2}\partial_{\mu}\overline\eta\partial^{\mu}\overline\eta+\dfrac{1+\delta^{\eta'}_k}{2}\partial_{\mu}\overline{\eta}'\partial^{\mu}\overline{\eta}'
+\delta_{k}^{\eta\eta'}\partial_{\mu}\overline\eta\partial^{\mu}\overline{\eta}'-\dfrac{m^{2}_{{\overline\eta}}+\delta_{m^{2}_{{\eta}}}}{2}\overline\eta\,\overline\eta-\dfrac{m^{2}_{{\overline{\eta}'}}+\delta_{m^{2}_{{\eta'}}}}{2}\overline{\eta}'\,\overline{\eta}'-\delta_{m^{2}}^{\eta\eta'}\overline\eta\,\overline{\eta}' \\
&+\dfrac{1+\delta^{\pi}_k}{2}\partial_{\mu}\overline{\pi}^{0}\partial^{\mu}\overline{\pi}^{0}+\delta^{\pi\eta}_{k}\partial_{\mu}\overline{\pi}^{0}\partial^{\mu}\overline\eta+\delta^{\pi\eta'}_{k}
\partial_{\mu}\overline{\pi}^{0}\partial^{\mu}\overline{\eta}'-\dfrac{m_{\overline{\pi}}^{2}+\delta_{m^{2}_{\pi}}}{2}\overline{\pi}^{0}\,\overline{\pi}^{0}-\delta^{\pi\eta}_{m^{2}}\overline{\pi}^{0}\overline{\eta}-\delta^{\pi{\eta}'}_{m^{2}}\overline{\pi}^{0}\overline{\eta}'\\
&+\dfrac{1+\delta^{a}_k}{2}\partial_{\mu}\overline{a}\partial^{\mu}\overline{a}+\delta^{a\pi}_{k}\partial_{\mu}\overline{a}\partial^{\mu}\overline{\pi}^{0}+\delta^{a\eta}_{k}\partial_{\mu}\overline{a}
\partial^{\mu}\overline\eta+\delta^{a{\eta}'}_{k}\partial_{\mu}\overline{a}
\partial^{\mu}\overline{\eta}'-\dfrac{m^{2}_{\overline{a}}+\delta_{m^{2}_{a}}}{2}\overline{a}\,\overline{a}-\delta^{a\pi}_{m^{2}}\overline{a}\,\overline{\pi}^{0}\\
&-\delta^{a\eta}_{m^{2}}\overline{a}\,\overline{\eta}-\delta^{a{\eta}'}_{m^{2}}\overline{a}\,\overline{\eta}'\,,
\end{aligned}
\end{equation}
where the $\delta_k^{X}$ with subscript $k$ denotes the corrections from higher orders to the kinetic terms and other $\delta_{j}$ corresponds to the higher order contributions to the mass terms.
It is noted that the higher derivative terms are contributed by the $O(p^6)$ and higher-order operators~\cite{Guo:2015xva} and hence are absent at NLO. Comparing with the results of only focusing the $\eta$-$\eta'$ mixing in Ref.~\cite{Guo:2015xva,Gu:2018swy}, we extend the calculations by simultaneously including the $\pi^0$ and the axion $a$ in the above equation. All the $\delta_i$ parameters in Eq.~\eqref{eq.lagsenlo} can be calculated order by order in the $U(3)$ chiral perturbation theory. Their explicit expressions calculated from the NLO Lagrangians of Eq.~\eqref{eq.lagnlo} are given in the Appendix.

Aside from the mass mixing terms, one also has to deal with the kinetic mixing at NLO. This can be done in a two-step procedure~\cite{Guo:2015xva}. In the first step, not only one needs to  eliminate the kinetic mixing terms, but also one should make sure that all the fields are in the canonical normalization, i.e., the bilinear derivative term of each field is multiplied by 1/2. In the second step, the mass mixing will be handled by an orthogonal matrix that does not spoil the already accomplished diagonalization of the kinetic-energy term \cite{Jamin:2000wn,Jamin:2001zq}. We will stick to the $\delta$ expansion up to NLO in all of these procedures. The diagonalized canonical fields, which are labeled with hats on top of each state, can be obtained via the following field redefinitions 
\begin{align}\label{eq.mixnlo} 
\left( \begin{array}{c}
\hat{\pi}^0 \\   \hat{\eta}  \\  \hat{\eta}' \\ \hat{a} 
\end{array} \right) 
\,=\, &
\left( \begin{array}{cccc}
1   & -y_{12} & -y_{13} & -y_{14} \\ 
y_{12} & 1  & -y_{23} & -y_{24}  \\ 
y_{13} & y_{23} & 1  & -y_{34} \\ 
y_{14} & y_{24} & y_{34} & 1 
\end{array} \right)
\times 
\nonumber \\ & 
\left( \begin{array}{cccc}
1- x_{11} & -x_{12} & -x_{13} & -x_{14} \\ 
-x_{12} & 1- x_{22} & -x_{23} & -x_{24}  \\ 
-x_{13} & -x_{23} & 1-x_{33} & -x_{34} \\ 
-x_{14} & -x_{24} & -x_{34} & 1-x_{44} 
\end{array} \right) \,
\left( \begin{array}{c}
\overline{\pi}^0 \\  \overline{\eta}  \\  \overline{\eta}' \\ \overline{a}
\end{array} \right) \,,
\end{align}
where the $x_{ij}$ and $y_{ij}$ are introduced to deal with the kinetic and mass mixing terms, respectively. 
The matrix elements of $x_{ij}$ are found to be 
\begin{eqnarray}
&& x_{11}= -\frac{\delta^\pi_k}{2}\,,\quad x_{12}= -\frac{\delta_k^{\pi\eta}}{2}\,,\quad x_{13}= -\frac{\delta_k^{\pi\eta'}}{2}\,,\quad x_{14}= -\frac{\delta_k^{a\pi}}{2}\,, \quad x_{22}= -\frac{\delta^{\eta}_k}{2}\,, \nonumber \\ &&
x_{23}= -\frac{\delta_k^{\eta\eta'}}{2}\,,\quad x_{24}= -\frac{\delta_k^{a\eta}}{2}\,, \quad x_{33}= -\frac{\delta^{\eta'}_k}{2}\,,\quad x_{34}= -\frac{\delta_k^{a\eta'}}{2}\,,\quad x_{44}= -\frac{\delta^{a}_k}{2}\,, 
\end{eqnarray}
and the expressions of the $y_{ij}$ read  
\begin{eqnarray}\label{eq.yij}
&&  y_{12}= \frac{\delta_{m^2}^{\pi\eta}+x_{12}(m_{\overline{\eta}}^2+m_{\overline{\pi}}^2)}{m_{\overline{\eta}}^2-m_{\overline{\pi}}^2}\,, \quad  y_{13}= \frac{\delta_{m^2}^{\pi\eta'}+x_{13}(m_{\overline{\eta}'}^2+m_{\overline{\pi}}^2)}{m_{\overline{\eta}'}^2-m_{\overline{\pi}}^2}\,, \quad  y_{14}= \frac{\delta_{m^2}^{a\pi}+x_{14}(m_{a,0}^2+m_{\overline{\pi}}^2)}{m_{a,0}^2-m_{\overline{\pi}}^2}\,, \nonumber \\ &&
 y_{23}= \frac{\delta_{m^2}^{\eta\eta'}+x_{23}(m_{\overline{\eta}}^2+m_{\overline{\eta}'}^2)}{m_{\overline{\eta}'}^2-m_{\overline{\eta}}^2}\,,\quad y_{24}= \frac{\delta_{m^2}^{a\eta}+x_{24}(m_{\overline{\eta}}^2+m_{a,0}^2)}{m_{a,0}^2-m_{\overline{\eta}}^2}\,,\quad y_{34}= \frac{\delta_{m^2}^{a\eta'}+x_{34}(m_{\overline{\eta}'}^2+m_{a,0}^2)}{m_{a,0}^2-m_{\overline{\eta}'}^2}\,. \nonumber \\
\end{eqnarray}

By keeping the contributions up to NLO, the masses of the diagonalized canonical states take the form 
\begin{eqnarray} \label{eq.massnlopi}
m_{\hat{\pi}}^2 &=& m_{\overline{\pi}}^2 + \delta_{m_\pi^2} - m_{\overline{\pi}}^2 \delta^\pi_k\,,  \\  \label{eq.massnloeta}
m_{\hat{\eta}}^2 &=& m_{\overline{\eta}}^2 + \delta_{m_{\eta}^2} - m_{\overline{\eta}}^2 \delta^{\eta}_k\,,  \\ \label{eq.massnloetap}
m_{\hat{\eta}'}^2 &=& m_{\overline{\eta}'}^2 + \delta_{m_{\eta'}^2} - m_{\overline{\eta}'}^2 \delta^{\eta'}_k\,,  \\
m_{\hat{a}}^2 &=& m_{\overline{a}}^2 + \delta_{m_a^2} - m_{a,0}^2 \delta^a_k\,,  \label{eq.massnloax}
\end{eqnarray} 
and the kaon mass squared up to NLO is similarly given by 
\begin{equation}\label{eq.massnlok}
 m_{\hat{K}}^2 = m_{\overline{K}}^2 + \delta_{m_K^2}- m_{\overline{K}}^2 \delta^{K}_{k}\,. 
\end{equation}
Up to NLO the relations between the physical states (denoted by the hatted fields) and the LO diagonalized ones reduce to 
\begin{eqnarray}
 \left( \begin{array}{c}
\hat{\pi}^0 \\   \hat{\eta}  \\  \hat{\eta}' \\ \hat{a} 
\end{array} \right) 
\,=\, &&
\left( \begin{array}{cccc}
1-x_{11}   & -x_{12}-y_{12} & -x_{13}-y_{13} & -x_{14}-y_{14} \\ 
-x_{12}+y_{12} & 1-x_{22}  & -x_{23}-y_{23} & -x_{24}-y_{24}  \\ 
-x_{13}+y_{13} & -x_{23}+y_{23} & 1-x_{33}  & -x_{34}-y_{34} \\ 
-x_{14}+y_{14} & -x_{24}+y_{24} & -x_{34}+y_{34} & 1-x_{44} 
\end{array} \right)
\left( \begin{array}{c}
\overline{\pi}^0 \\  \overline{\eta}  \\  \overline{\eta}' \\ \overline{a}
\end{array} \right) \,.
\end{eqnarray}

To combine the LO relation~\eqref{eq.mixlo08}, the mixing matrix between the physical states and the  bare ones are given by  
\begin{eqnarray}\label{eq.mixnlof}
 \left( \begin{array}{c}
\hat{\pi}^0 \\   \hat{\eta}  \\  \hat{\eta}' \\ \hat{a} 
\end{array} \right) 
\,=\, 
\left( \begin{array}{cccc}
1 + z_{11}    & -v_{12} + z_{12}   & -v_{13} + z_{13} & -v_{14}+z_{14} \\ 
v_{12} + z_{21} & 1+ z_{22}   &   z_{23} & -v_{24}+z_{24}  \\ 
v_{13} + z_{31} &  z_{32}  & 1+ z_{33}   & -v_{34}+z_{34}  \\ 
v_{14}+z_{41} & v_{24} +z_{42} &  v_{34}+z_{43} & 1+ z_{44} 
\end{array} \right)
\left( \begin{array}{c}
\pi^0 \\  \mathring{\overline{\eta}}  \\  \mathring{\overline{\eta}}' \\ a
\end{array} \right) \,,
\end{eqnarray}
where the NLO corrections are collected in the $z_{ij}$ and their explicit forms are given by 
\begin{eqnarray}
\label{221101.1}
z_{11}=&& -x_{11}\,, \nonumber \\ 
z_{12}=&& -x_{12}-y_{12}-v_{12}x_{22}+v_{13}(y_{23}-x_{23}) \,, \nonumber \\
z_{13}=&& -x_{13}-y_{13}-v_{13}x_{33}-v_{12}(y_{23}+x_{23}) \,, \nonumber\\
z_{14}=&& -x_{14}-y_{14}-v_{12}(x_{24}+y_{24}) - v_{13}(x_{34}+y_{34}) \,, \nonumber\\
z_{21}=&& -x_{12}+y_{12}+ v_{12}x_{11} \,, \nonumber\\
z_{22}=&& -x_{22}\,,\nonumber\\
z_{23}=&& -x_{23}-y_{23}\,, \nonumber\\ 
z_{24}=&& -x_{24}-y_{24}\,, \nonumber\\ 
z_{31}=&& -x_{13}+y_{13}+v_{13}x_{11}\,,\nonumber\\ 
z_{32}=&& -x_{23}+y_{23} \,, \nonumber\\ 
z_{33}=&& -x_{33}\,,\nonumber\\
z_{34}=&& -x_{34}-y_{34} \,, \nonumber\\ 
z_{41}=&& -x_{14}+y_{14}+v_{14}x_{11}+v_{24}(x_{12}-y_{12})+v_{34}(x_{13}-y_{13})\,, \nonumber\\ 
z_{42}=&& -x_{24}+y_{24}+v_{24}x_{22}+v_{34}(x_{23}-y_{23}) \,, \nonumber\\ 
z_{43}=&& -x_{34}+y_{34}+v_{34}x_{33}+v_{24}(x_{23}+y_{23}) \,, \nonumber\\ 
z_{44}=&& -x_{44}\,.
\end{eqnarray}

As a result of combining Eqs.~\eqref{eq.mixnlof} and \eqref{eq.lomixing}, one can obtain 
\begin{eqnarray}\label{eq.mixnlof08}
 {\tiny  \left( \begin{array}{c}
\hat{\pi}^0 \\   \hat{\eta}  \\  \hat{\eta}' \\ \hat{a} 
\end{array} \right) 
=
\left( \begin{array}{cccc}
1 + z_{11}      & c_\theta(-v_{12}+z_{12}) +s_\theta(-v_{13} + z_{13}) & -s_\theta(-v_{12}+z_{12}) +c_\theta(-v_{13} + z_{13}) & -v_{14}+z_{14} \\ 
v_{12} + z_{21} & c_\theta(1+ z_{22}) +s_\theta  z_{23}  & -s_\theta(1+ z_{22}) +c_\theta z_{23}    & -v_{24}+z_{24}  \\ 
v_{13} + z_{31} & c_\theta z_{32} +s_\theta(1+ z_{33})   & -s_\theta z_{32}+ c_\theta(1+ z_{33})    & -v_{34}+z_{34}  \\ 
v_{14}+z_{41}   & c_\theta(v_{24}+z_{42}) +s_\theta(v_{34}+z_{43})   & -s_\theta(v_{24}+z_{42})+ c_\theta(v_{34}+z_{43}) & 1+ z_{44} 
\end{array} \right)  
\left( \begin{array}{c}
\pi^0 \\  \eta_8 \\  \eta_0 \\ a
\end{array} \right) \,. }  
\end{eqnarray}
Alternatively one could also use the quark-flavor bases of $\eta_q$ and $\eta_s$, and they relate with the octet-singlet $\eta_8, \eta_0$ bases via 
\begin{eqnarray}\label{eq.mix08qs}
\left(\begin{array}{c} \eta_8 \\ \eta_0\end{array}\right)  =
\left(\begin{array}{cc}  \sqrt{\frac{1}{3}} & - \sqrt{\frac{2}{3}} \\  \sqrt{\frac{2}{3}} & \sqrt{\frac{1}{3}} \end{array}\right)  
\left(\begin{array}{c} \eta_q \\ \eta_s \end{array}\right)\,,
\end{eqnarray}
where the constituent quark contents of $\eta_q$ and $\eta_s$ are $(\bar{u}u+\bar{d}d)/\sqrt{2}$ and $\bar{s}s$, respectively. The LO diagonalized $\mathring{\overline{\eta}}, \mathring{\overline{\eta}}'$ states can be decomposed in terms of the quark-flavor base 
\begin{eqnarray}\label{eq.mixqs}
\left(\begin{array}{c} \mathring{\overline{\eta}} \\  \mathring{\overline{\eta}}' \end{array}\right)  =
\left(\begin{array}{cc}  \cos{\phi_{qs}} & -\sin{\phi_{qs}} \\  \sin{\phi_{qs}} & \cos{\phi_{qs}} \end{array}\right)  
\left(\begin{array}{c} \eta_q \\ \eta_s \end{array}\right)\,,
\end{eqnarray}
with $\phi_{qs}= \theta_{\rm id} + \theta$ and the ideal mixing angle $\theta_{\rm id}=\arcsin({\sqrt{2}/\sqrt{3}})$. Therefore, by combining Eqs.~\eqref{eq.mixnlof} and  \eqref{eq.mixqs}, one can in turn obtain the relations between the physical states and those in the quark-flavor bases. The results share the same forms as those in Eq.~\eqref{eq.mixnlof08} with the explicit replacement of the angle $\theta$ by the angle $\phi_{qs}$ defined in Eq.~\eqref{eq.mixqs}. 

The two-angle mixing formula proposed in Ref.~\cite{Leutwyler:1997yr} gives an intuitive relation  between the physical states denoted by $\hat{\eta}^{(')}$ here and the ones in octet-singlet basis   
\begin{eqnarray} \label{eq.twoanglesmixing08}
 \left(
 \begin{array}{c}
 \hat{\eta}   \\
 \hat{\eta}' \\
 \end{array}
 \right) = \frac{1}{F}\left(
                                        \begin{array}{cc}
                                          F_8\, \cos{\theta_8}  & -F_0 \,\sin{\theta_0}  \\
                                           F_8\,\sin{\theta_8} & F_0 \,\cos{\theta_0} \\
                                        \end{array}
    \right)
      \left(
       \begin{array}{c}
       \eta_8   \\
       \eta_0  \\
       \end{array}
        \right)\,,
\end{eqnarray}
which naturally reduces to the conventional mixing relation with one angle by taking $F_8=F_0=F$ and $\theta_0=\theta_8=\theta$.  
Similarly one could also introduce the two-angle mixing formalism to the quark-flavor base
\begin{eqnarray} \label{eq.twoanglesmixingqs}
 \left(
 \begin{array}{c}
 \hat{\eta}   \\
 \hat{\eta}' \\
 \end{array}
 \right) = \frac{1}{F}\left(
                                        \begin{array}{cc}
                                          F_q\, \cos{\theta_q}  & -F_s \,\sin{\theta_s}  \\
                                           F_q\,\sin{\theta_q} & F_s \,\cos{\theta_s} \\
                                        \end{array}
    \right)
      \left(
       \begin{array}{c}
       \eta_q   \\
       \eta_s  \\
       \end{array}
        \right)\,.
\end{eqnarray}

\begin{figure}[htbp]
\centering
\includegraphics[width=0.49\textwidth,angle=-0]{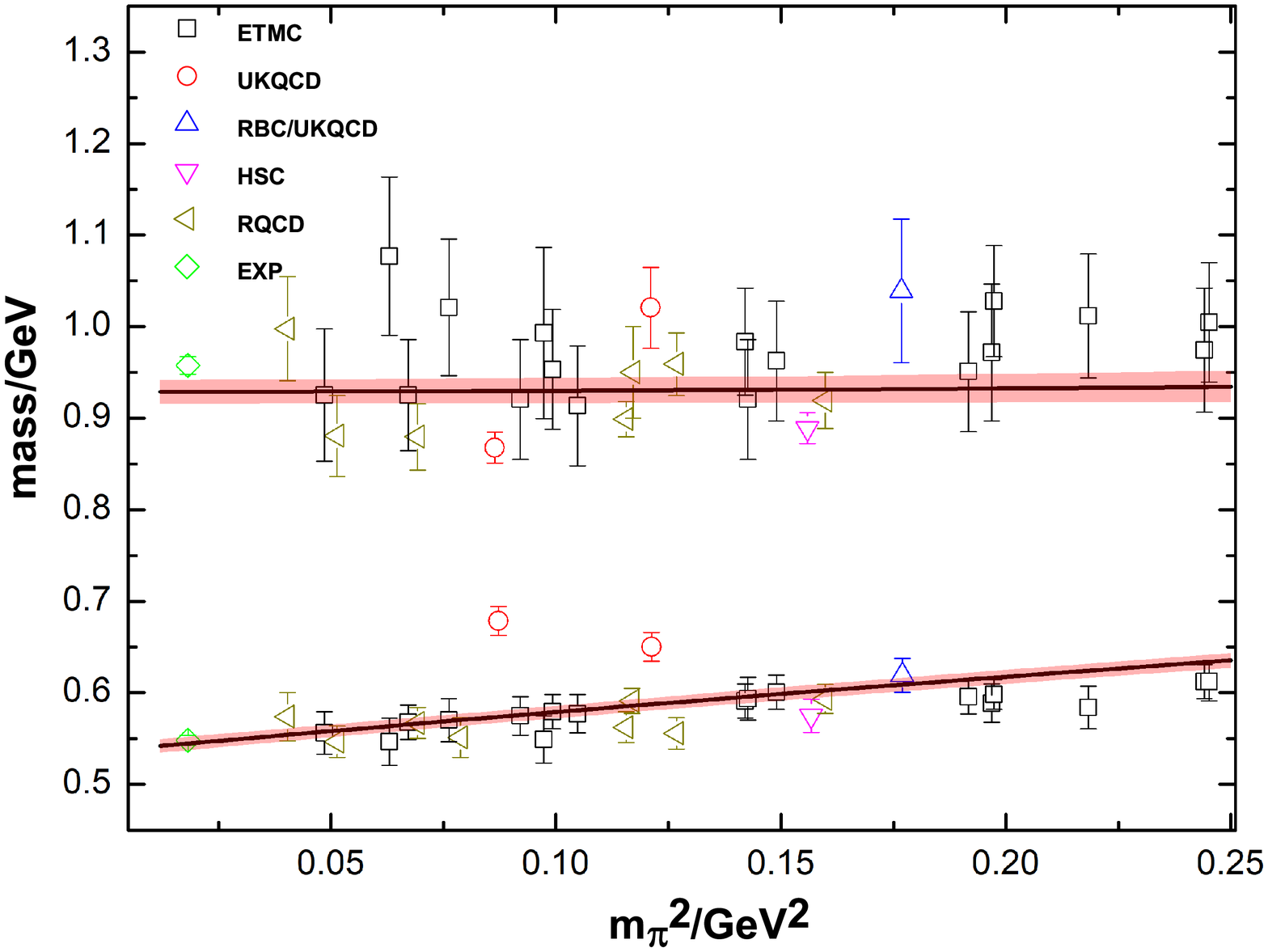} 
\includegraphics[width=0.49\textwidth,angle=0]{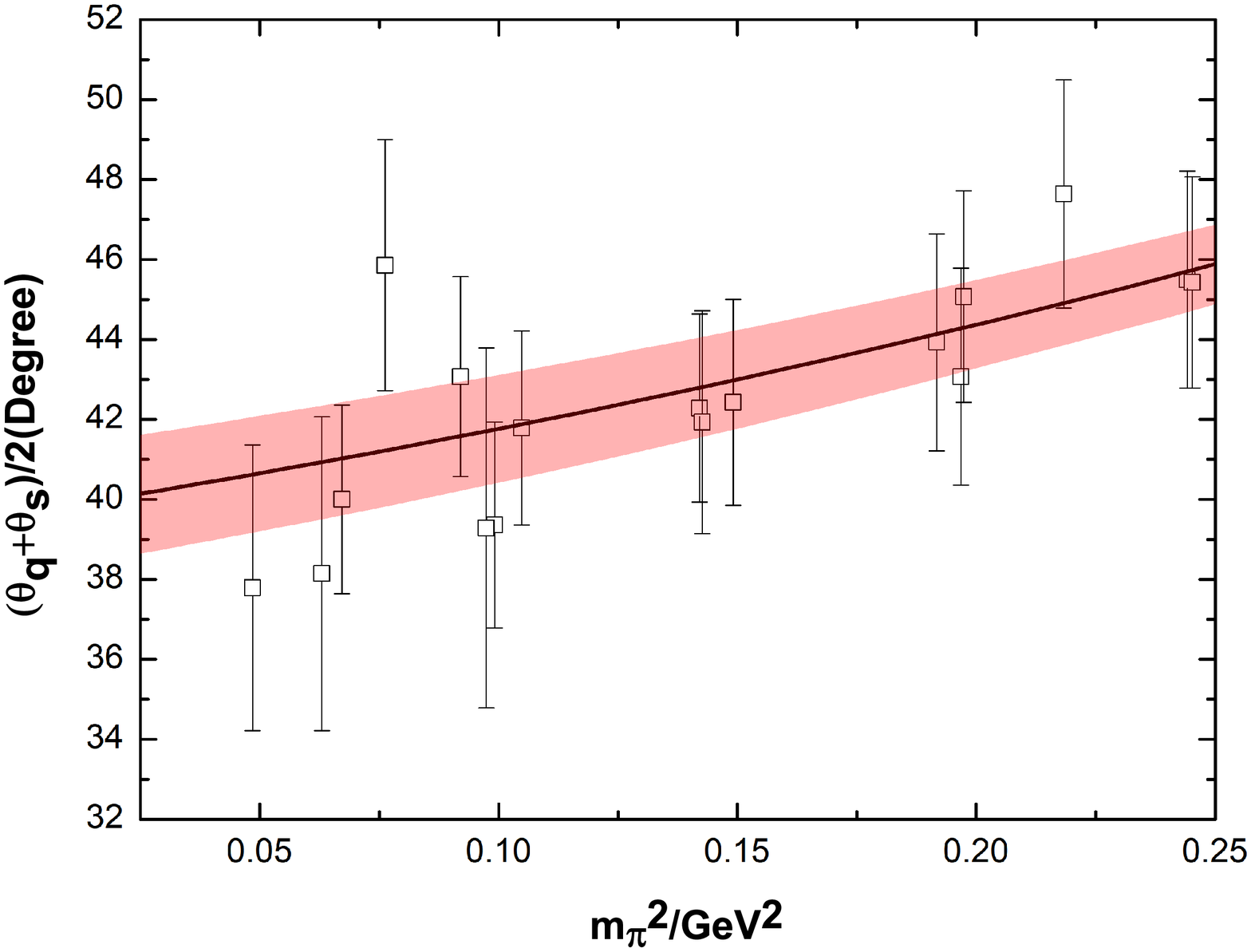} 
\caption{ The masses of $\eta$ and $\eta'$ as functions of $m_\pi$ (left panel) and the pion-mass dependences of the $\eta$-$\eta'$ mixing angle in the quark-flavor basis (right panel). The lattice data of the $\eta/\eta'$ masses are taken from  ETMC~\cite{Ottnad:2017bjt}, UKQCD~\cite{Gregory:2011sg}, RBC/UKQCD~\cite{Christ:2010dd}, HSC~\cite{Dudek:2011tt} and RQCD~\cite{Bali:2021qem}. For the $\eta/\eta'$ masses from Ref.~\cite{Bali:2021qem}, only the results from the ensemble with approximately physical mass of strange quark are included. The lattice data of mixing angles in the quark-flavor base are taken from Ref.~\cite{Ottnad:2017bjt}.  } \label{fig.etaetap}
\end{figure}

\begin{figure}[htbp]
\centering
\includegraphics[width=0.49\textwidth,angle=-0]{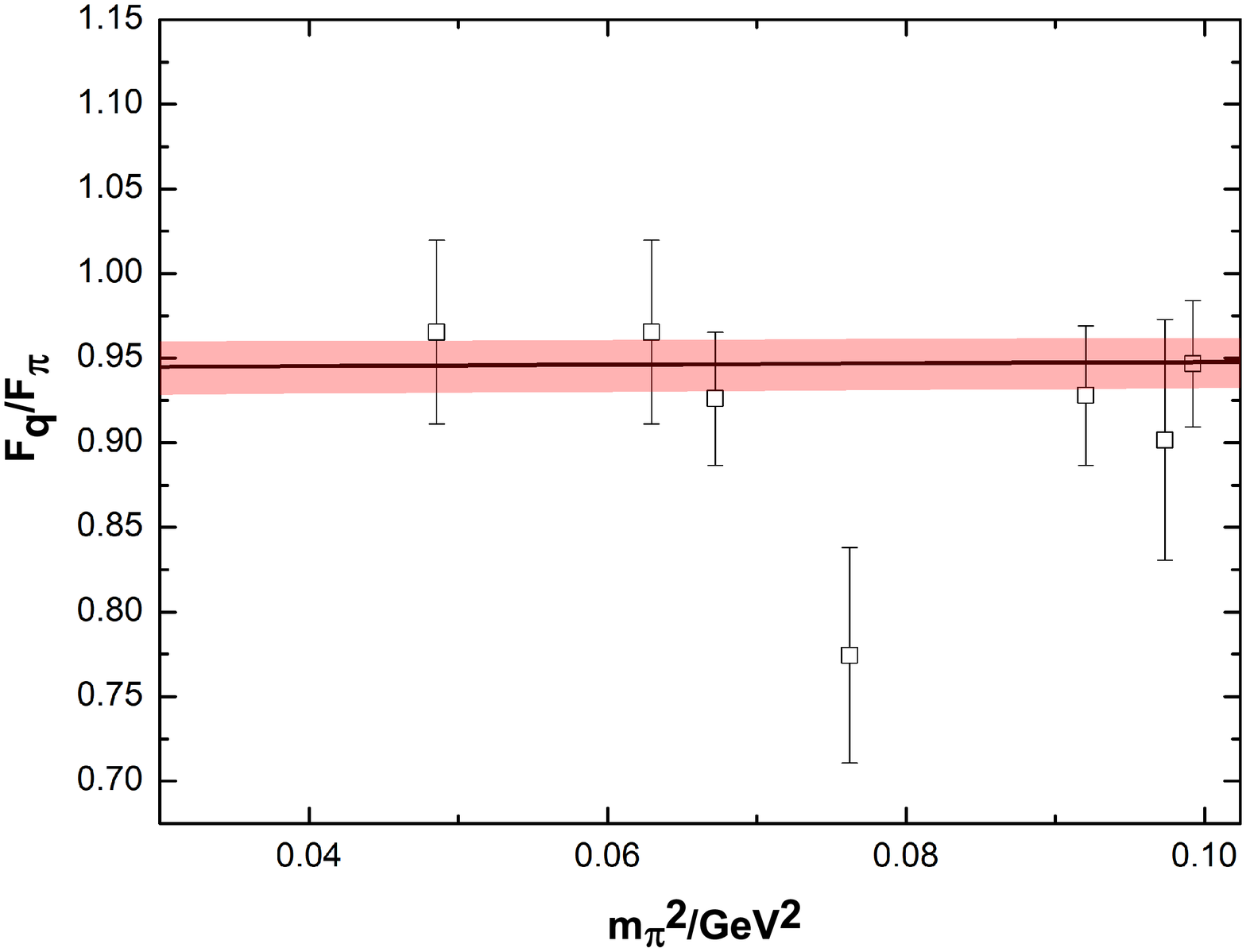} 
\includegraphics[width=0.49\textwidth,angle=-0]{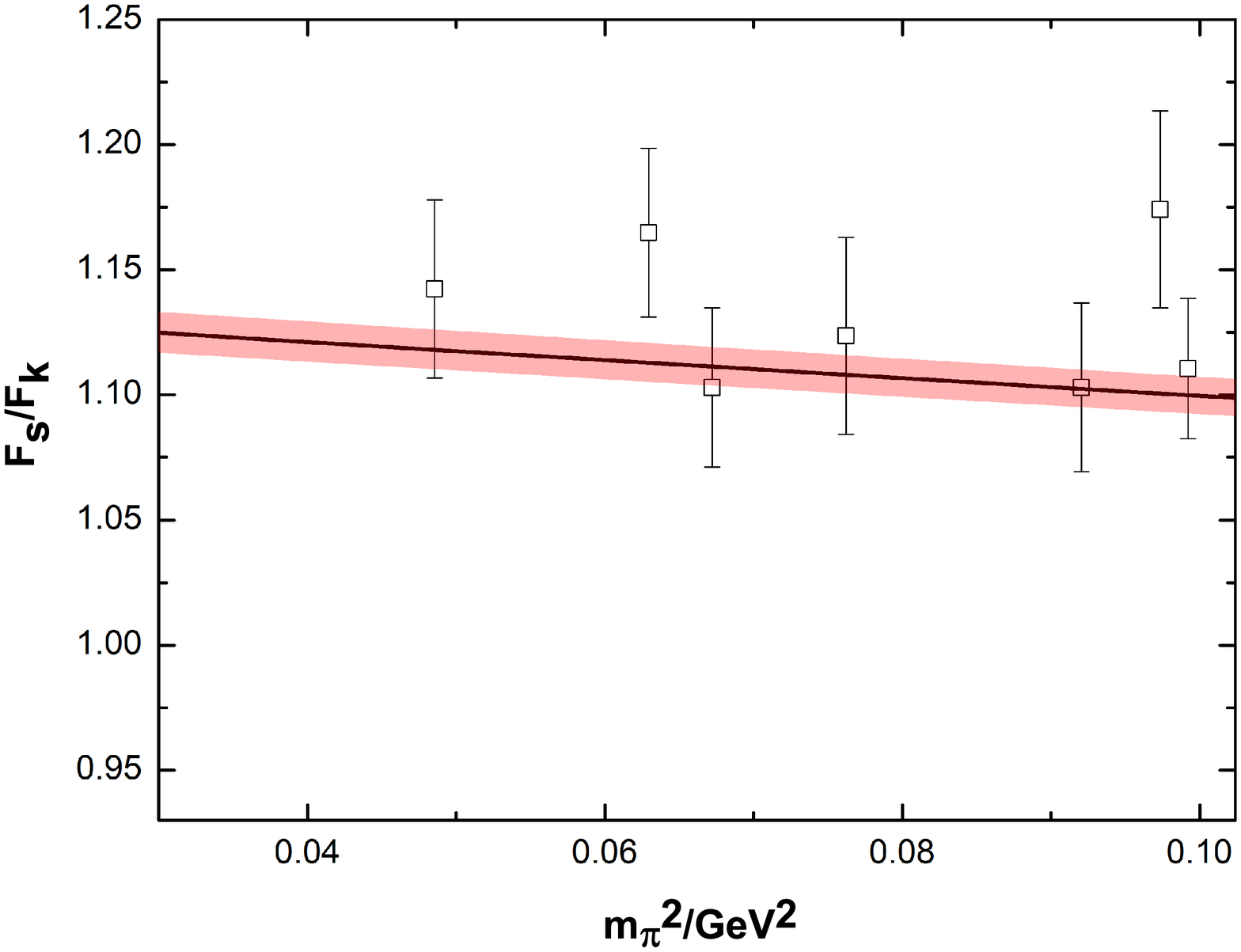} 
\caption{ The decay constants of $F_q$ and $F_s$ defined in Eq.~\eqref{eq.twoanglesmixingqs} as functions of $m_\pi$. The lattice data are taken from Ref.~\cite{Ottnad:2017bjt}.   } \label{fig.fqfs}
\end{figure}

\begin{figure}[htbp]
\centering
\includegraphics[width=0.32\textwidth,angle=-0]{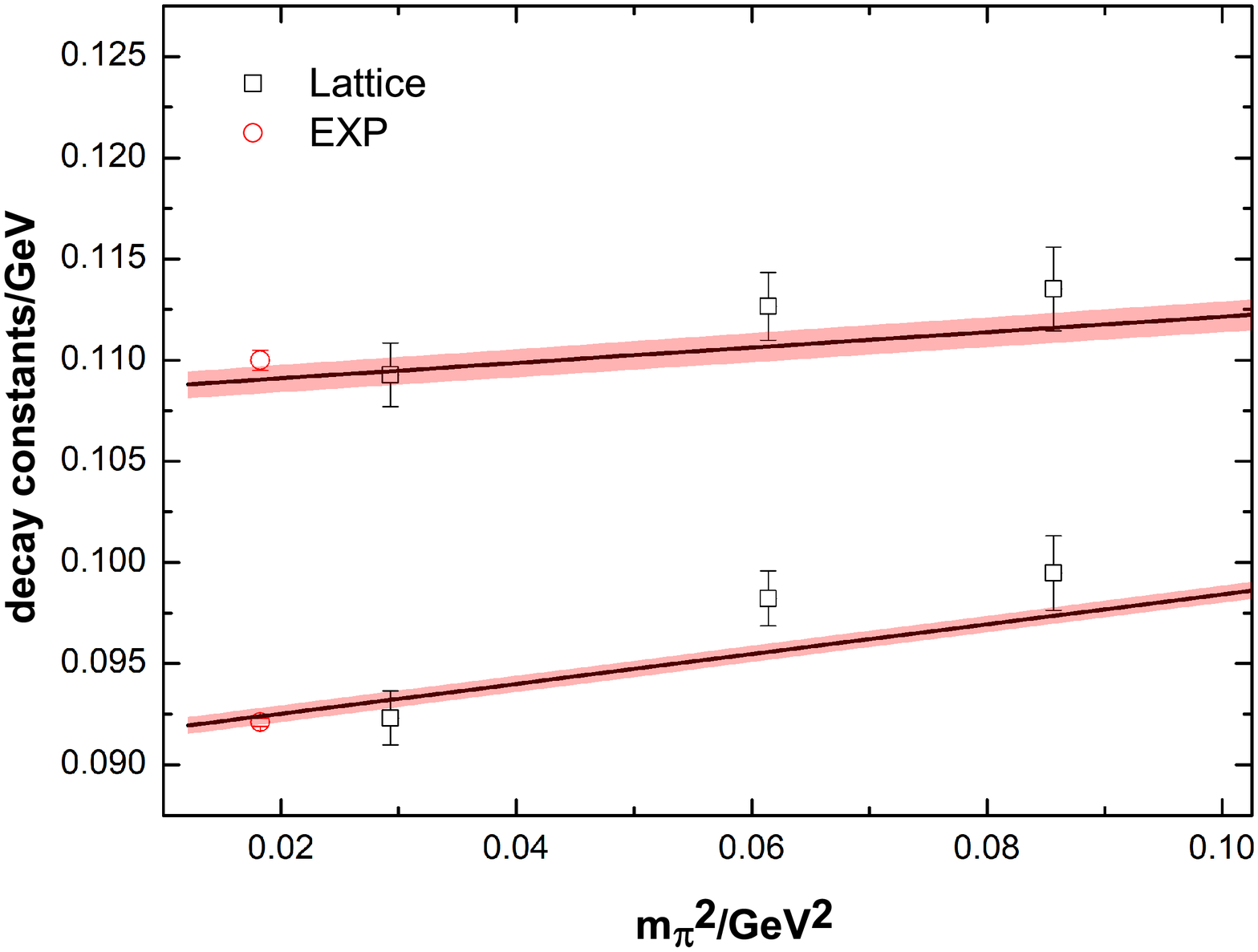} 
\includegraphics[width=0.32\textwidth,angle=-0]{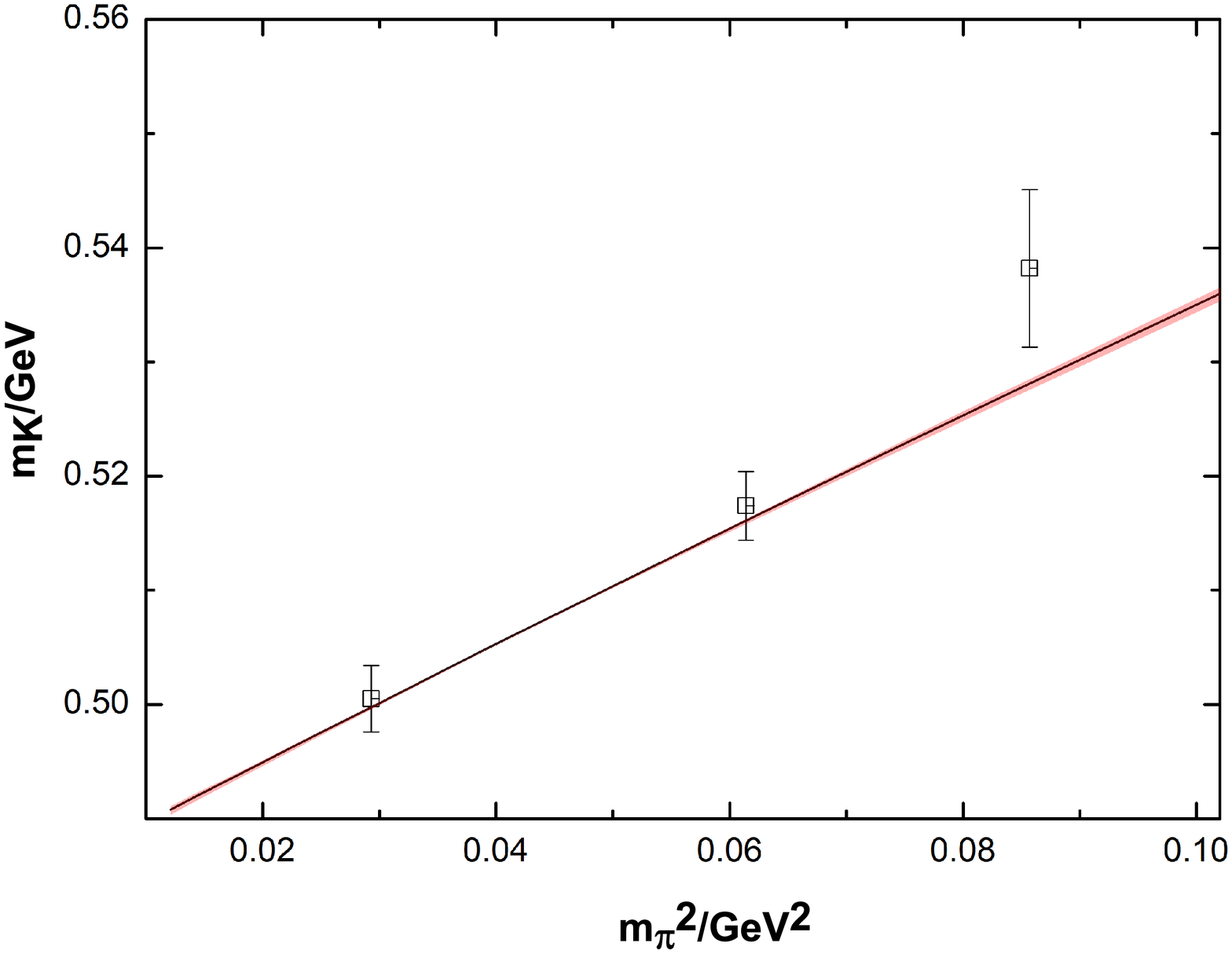} 
\includegraphics[width=0.32\textwidth,angle=-0]{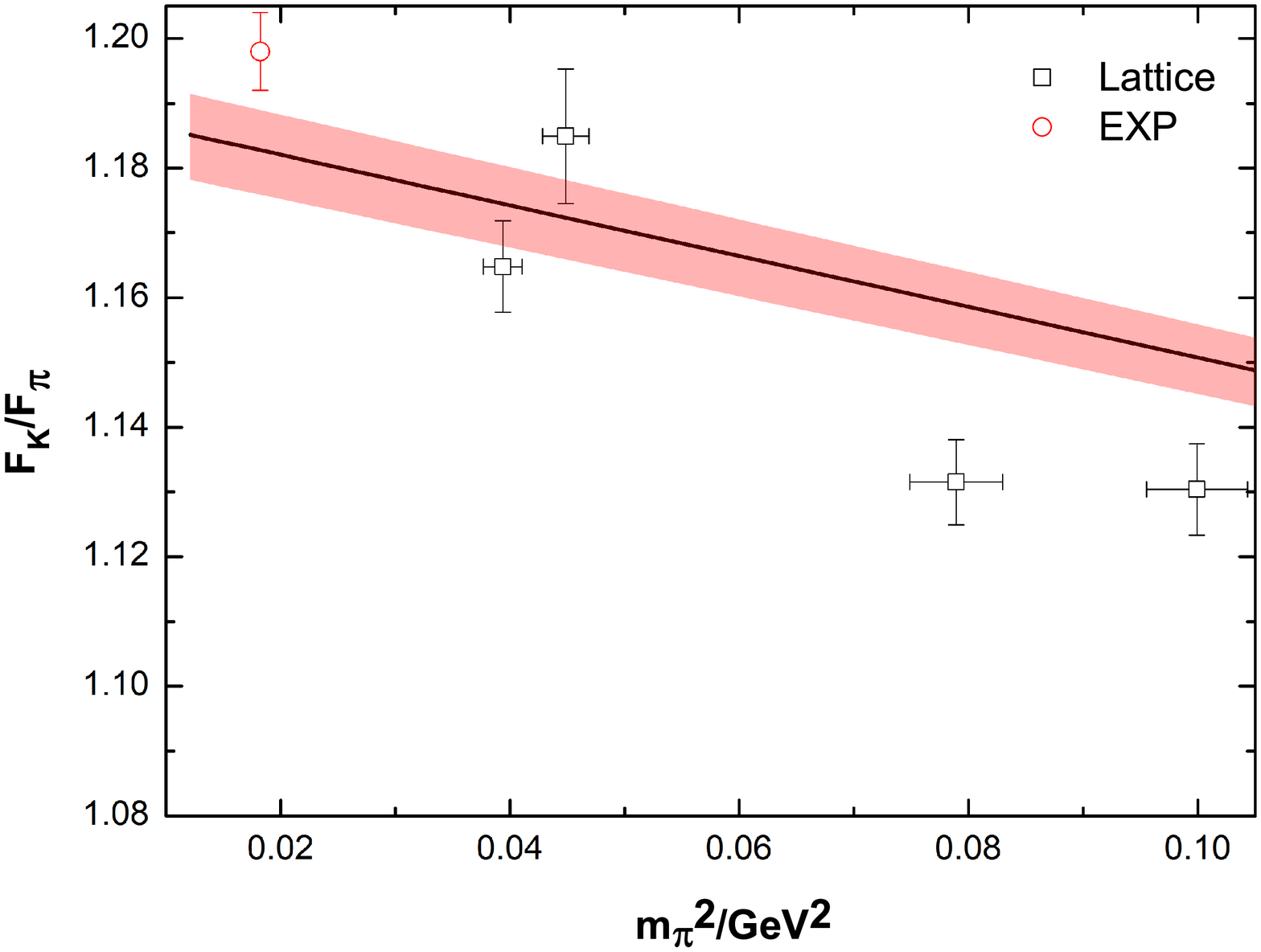} 
\caption{ The pion-mass dependences of $F_\pi$, $F_K$ and $m_K$. The lattice data in the left and central panels are from Refs.~\cite{RBC:2010qam,RBC:2012cbl}. The data in the right panel are from Ref.~\cite{Durr:2010hr}. } \label{fig.fpifkmk}
\end{figure}

Remarkable progress has been made in recent years for the lattice calculation of the $\eta$-$\eta'$ system. Both their masses and mixing angles at different pion masses are obtained  by many lattice collaborations. 
In this work, we fix the unknown low-energy constants (LECs) of $\xpt$ by performing fits to relevant lattice data, including the $\eta$/$\eta'$ masses and their mixing angle at different $m_\pi$, the weak decay constants of pion and kaon, and the pion-mass dependences of $m_K$. Apart from the lattice data of the $\eta$/$\eta'$ masses at unphysically large $m_\pi$ from the ETMC~\cite{Ottnad:2017bjt}, UKQCD~\cite{Gregory:2011sg}, RBC/UKQCD~\cite{Christ:2010dd}, HSC~\cite{Dudek:2011tt} that have been analyzed in Refs.~\cite{Guo:2015xva,Gu:2018swy}, we also include the new results from the RQCD collaboration~\cite{Bali:2021qem} in the present study. In order to make direct comparisons of other data, only the lattice simulations with physical strange quark mass from Ref.~\cite{Bali:2021qem} are taken into account here. The various lattice data of the masses for $\eta$ and $\eta'$ are shown in the left panel of Fig.~\ref{fig.etaetap}, together with their mixing angles in the quark-flavor base in the right panel. To be in accordance with the lattice setup, we take the average of $\theta_q$ and $\theta_s$ to fit the latter data set. It is noted that the phenomenological determinations also favor rather similar values for $\theta_q$ and $\theta_s$~\cite{Chen:2012vw,Chen:2014yta,Guo:2015xva}. 
The decay constants $F_q$ and $F_s$ defined in the mixing matrix of the quark-flavor base as functions of pion masses are provided by the lattice study~\cite{Ottnad:2017bjt} and we will take these kinds of data in our fits too, as shown in Fig.~\ref{fig.fqfs}. 
For the pion-mass dependences of the $F_\pi, F_K$ and $m_K$, we take into account the lattice data up to $m_\pi^2=0.1$~GeV$^2$, as explicitly illustrated in Fig.~\ref{fig.fpifkmk}.

\begin{table}[htbp]
 \centering
\begin{tabular}{l l}
\hline\hline
  Parameters                  & NLO Fit 
            \\\hline
$F({\rm MeV})$        &$91.05^{+0.42}_{-0.44}$   \\  
$10^3\times L_5$      &$1.68^{+0.05}_{-0.06}$ \\    
$10^3\times L_8$      &$0.88^{+0.04}_{-0.04}$  \\    
$\Lambda_1$           &$-0.17^{+0.05}_{-0.05}$  \\   
$\Lambda_2$           &$0.06^{+0.08}_{-0.09}$  \\    
$\chi^2/(d.o.f)$      &$219.9/(111-5)$ \\  
\hline\hline
\end{tabular}
\caption{\label{tab.fitnlo} The values of the LECs from the NLO fit. } 
\end{table}

The  values of the LECs from the fits are summarized in Table~\ref{tab.fitnlo}. The resulting curves for the masses of $\eta$ and $\eta'$ and their mixing angles from the fits are given in Fig.~\ref{fig.etaetap}.  It is noted that the LO fit to the masses of $\eta$ and $\eta'$ with just the single parameter $M_0$ can reasonably reproduce the lattice data with $M_0=820$~MeV~\cite{Guo:2015xva,Gu:2018swy}. Therefore we will fix $M_0$ at this value during the NLO fits. It is verified that by releasing the $M_0$ the fits do not improve and the values of the parameters in Table~\ref{tab.fitnlo} barely change. The resulting parameters from the revised NLO fits are close to the previous ones given in Refs.~\cite{Guo:2015xva,Gu:2018swy}. With  the fitted parameters in Table~\ref{tab.fitnlo}, we are ready to predict the important quantities related with the axion. 

Since the bare mass of the axion, i.e. $m_{a,0}$ in Eq.~\eqref{eq.lagag}, is explicitly kept throughout our calculations, e.g.  Eqs.~\eqref{eq.massnloax},~\eqref{eq.v12}-\eqref{eq.v34} and \eqref{eq.yij}, it is  straightforward for us to explore the so-called axion-like scenarios by assigning some specific nonvanishing values to $m_{a,0}$. Nevertheless, in this work we will mainly focus on the phenomenological predictions for the QCD-axion scenario by taking $m_{a,0}=0$ in Eq.~\eqref{eq.lagag}.  
The explicit values of the transition matrix elements in Eq.~\eqref{eq.mixnlof08} between the physical states of $\hat{\pi}^0, \hat{\eta}, \hat{\eta}', \hat{a}$ and the bare states of $\pi^0, \eta_8, \eta_0, a$ are determined to be 
\begin{eqnarray}\label{eq.mixnlof08numdef}
  \left( \begin{array}{c}
\hat{\pi}^0 \\   \hat{\eta}  \\  \hat{\eta}' \\   \hat{a} 
\end{array} \right) 
= M^{\rm LO+NLO} 
\left( \begin{array}{c}
\pi^0 \\  \eta_8 \\  \eta_0 \\ a 
\end{array} \right) \,,   
\end{eqnarray}
with 
\begin{eqnarray}\label{eq.mixnlof08num}
&& M^{\rm LO+NLO} 
= \nonumber \\ &&
{\tiny \left( \begin{array}{cccc}
1+ (0.015\pm 0.001)     & 0.017 +(-0.010\pm 0.001) & 0.009+ (-0.007\pm 0.001)  & \frac{12.1+ (0.48\pm 0.08)}{f_a} \\ 
-0.019+(0.007\pm 0.001) & 0.94 +(0.21\pm 0.01) & 0.33+(-0.22\pm 0.03) & \frac{34.3+ (0.9\pm 0.2)}{f_a}  \\ 
-0.003+(-0.003\pm 0.000) & -0.33 +(-0.18\pm 0.03) &  0.94+(0.13\pm 0.02)  & \frac{25.9+(-0.5\pm 0.1)}{f_a}  \\ 
\frac{-12.1+(-0.20\pm 0.03)}{f_a} & \frac{-23.8+ (1.6^{+0.8}_{-0.8})}{f_a}  & \frac{-35.7+(-5.7^{+1.6}_{-1.7})}{f_a} & 1 + \frac{27.6\pm 1.0}{f_a^2} 
\end{array} \right)\,,}  \nonumber \\ && 
\end{eqnarray}
where the first and second entries in each matrix element correspond to the LO and NLO contributions, respectively. For the numbers in the last row and last column accompanying $1/f_a$, they are given in units of MeV, while the number accompanying $1/f_a^2$ is given in units of MeV$^2$. The main focus of our work is the relative corrections from the NLO part compared to the LO one.

For the mixing strengths between the $\hat{\pi}^0$ and $\eta_{0,8}$, and also the mixing between the $\hat{\eta}^{(')}$ and the $\pi^0$, which are all proportional to the leading IB factor $\epsilon$, the NLO parts in the $\delta$ counting gives rather large relative corrections compared to the LO results, ranging from around 60\% up to 100\%. For the $\eta$-$\eta'$ mixing, it is also found that the NLO contribution can be as large as around 60\% relative to the LO part. However, it is interesting to note that all the mixing strengths between the axion and the light-flavor pseudoscalar $\pi^0, \eta, \eta'$, i.e. the numbers in the last column and the last row, the relative NLO corrections in the $\delta$ counting to the LO parts are small, ranging from around 2\% up to around 15\%. The uncertainties of the mixing strengths given in Eq.~\eqref{eq.mixnlof08num}, estimated by the using the error bands of the LECs in Table~\ref{tab.fitnlo}, turn out to be mild as well.

The contributions from the LO and NLO parts to the mass squared of the light-flavor pseudoscalar mesons and the axion in Eqs.~\eqref{eq.massnlopi}-\eqref{eq.massnlok}  are found to be 
\begin{eqnarray}
m_{\hat\pi}&=& \big[ 134.90 + (0.10 \pm 0.07 ) \big] {\rm MeV}  \,, \nn\\
m_{\hat K}&=& \big[ 489.2 + (5.0^{+3.4}_{-3.5} ) \big] {\rm MeV}  \,, \nn\\
m_{\hat\eta}&=& \big[ 490.2 + (60.9^{+10.2}_{-10.0})  \big] {\rm MeV} \,, \nn\\
m_{\hat{\eta}'} &=& \big[ 954.3+(-28.4^{+11.9}_{-12.6})  \big] {\rm MeV}\,, \nn\\
m_{\hat{a}} &=& \big[5.96 + (0.12\pm 0.02) \big] \mu {\rm eV} \frac{10^{12} {\rm GeV}}{f_a}\,,  
\end{eqnarray}
where the first and second entries inside the square brackets denote the results from the LO and NLO terms, respectively. Comparing with the LO case, it is clear that the NLO correction brings the $\eta$ mass much closer to its physical value. When compared with the LO case, the NLO correction brings the $\eta$ mass much closer to its physical value, although it somewhat worsens the description of the $\eta'$ mass. However, we should notice that while the NLO contribution to the $\eta'$ mass is quite small, less than 3\% of the LO value, it is a 12\% for  the $\eta$ mass, which allows us to match its experimental value. The NLO contribution to the axion mass turns out to be rather small. 
 
\section{Two-photon couplings }\label{sec.twophoton}

Relying on the previous results of the mixing relations, we are ready to study the two-photon decays of the $\pi^0,\eta,\eta'$ and $a$, based on the LO and NLO WZW operators in Eqs.~\eqref{eq.lagwzwlo} and \eqref{eq.lagwzwnloexpd}, respectively. By inserting the mixing relations~\eqref{eq.mixnlof} into the LO WZW Lagrangian~\eqref{eq.lagwzwlo} and neglecting all the IB terms, the two-photon couplings of the physical states read 
\begin{eqnarray}
\mathcal{L}_{WZW}^{\rm LO}=&&  -\frac{3\sqrt{2}e^2}{8\pi^2 F}\varepsilon_{\mu\nu\rho\sigma}\partial^\mu A^\nu \partial^\rho A^\sigma \bigg\{ \frac{1+x_{11}}{3\sqrt{2}}\hat{\pi}^0   \nonumber\\&& 
+\bigg[ \frac{c_\theta -2\sqrt{2}s_\theta}{3\sqrt{6}} (1+x_{22}) +\frac{s_\theta +2\sqrt{2}c_\theta}{3\sqrt{6}} (x_{23}-y_{23}) \bigg]  \hat{\eta} + \nonumber\\ && 
+ \bigg[ \frac{s_\theta +2\sqrt{2}c_\theta}{3\sqrt{6}} (1+x_{33}) +\frac{c_\theta -2\sqrt{2}s_\theta}{3\sqrt{6}} (x_{23}+y_{23}) \bigg] \hat{\eta}' +  \nonumber\\&&
+ \bigg[  \frac{(c_\theta -2\sqrt{2}s_\theta)v_{24}+ (s_\theta +2\sqrt{2}c_\theta)v_{34}}{3\sqrt{6}} 
\nonumber \\&& 
\qquad +\frac{(c_\theta -2\sqrt{2}s_\theta)(x_{24}+y_{24})+(s_\theta +2\sqrt{2}c_\theta)(x_{34}+y_{34})}{3\sqrt{6}}  \bigg]\hat{a} \bigg\} \,,
\end{eqnarray}
where all of the NLO terms are introduced through the mixing. Since we stick to the NLO calculation, only the LO mixing relations~\eqref{eq.mixlo08} are needed to insert into the NLO WZW Lagrangian~\eqref{eq.lagwzwnloexpd} and the resulting two-photon interactions are given by 
\begin{eqnarray}
 \mathcal{L}_{WZW}^{\rm NLO}=&& \frac{e^2}{F}\varepsilon_{\mu\nu\rho\sigma}\partial^\mu A^\nu \partial^\rho A^\sigma \bigg\{ \frac{32m_\pi^2 t_1}{3}\hat{\pi}^0  \nonumber \\&&
+\frac{32}{9\sqrt{3}} \big\{ -9\sqrt{2} s_\theta k_3 + t_1 \big[ c_\theta (7m_\pi^2-4m_K^2) -2\sqrt{2}s_\theta(m_K^2+2m_\pi^2)  \big] \big\}\hat{\eta} 
 \nonumber \\&&
+\frac{32}{9\sqrt{3}} \big\{ 9\sqrt{2} c_\theta k_3 + t_1 \big[ 2\sqrt{2}c_\theta (2m_\pi^2+m_K^2) +s_\theta(7m_\pi^2-4m_K^2)  \big] \big\}\hat{\eta}' 
\nonumber \\&&
+\frac{32}{27} \big\{ 9 k_3\big(\frac{F}{f_a}-\sqrt{6}s_\theta v_{24}+\sqrt{6}c_\theta v_{34} \big) 
\nonumber \\ && 
\qquad\quad \,\,  + t_1 \big[ -\sqrt{3}s_\theta(4\sqrt{2}v_{24}m_\pi^2-7v_{34}m_\pi^2+  2\sqrt{2}v_{24}m_K^2+4v_{34}m_K^2) 
\nonumber \\ &&
\qquad \qquad \qquad +\sqrt{3}c_\theta(4\sqrt{2}v_{34}m_\pi^2+7v_{24}m_\pi^2+  2\sqrt{2}v_{34}m_K^2-4v_{34}m_K^2)    \big] \big\}\hat{a} 
\bigg\} \,. \nn \\
\end{eqnarray}

The two-photon decay amplitude of $\phi\to\gamma(k_1)\gamma(k_2)$ with $\phi=\pi^0,\eta,\eta'$ and $a$ can be written as 
\begin{eqnarray}\label{eq.defTphigg}
 T_{\phi\to \gamma\gamma} = e^2\varepsilon_{\mu\nu\rho\sigma} k_1^\mu \epsilon_1^\nu k_2^\rho \epsilon_2^{\sigma} F_{\phi\gamma\gamma}\,,
\end{eqnarray}
where the two-photon coupling strengths $F_{\phi\gamma\gamma}$ read 
\begin{eqnarray}\label{eq.fpigg}
 F_{\pi^0\gamma\gamma} = \frac{1}{4\pi^2 F} +\frac{1}{4\pi^2 F}x_{11} - \frac{64}{3 F}t_1 m_\pi^2 \,,
\end{eqnarray}
\begin{eqnarray}\label{eq.fetagg}
F_{\eta\gamma\gamma} =&& \frac{c_\theta -2\sqrt{2}s_\theta}{4\sqrt{3}\pi^2 F} (1+x_{22})+ \frac{s_\theta +2\sqrt{2}c_\theta}{4\sqrt{3}\pi^2F} (x_{23}-y_{23}) + \frac{64\sqrt{6}}{3F} s_\theta k_3 
\nonumber \\ && 
-  \frac{64\sqrt{3}}{27F}t_1 \bigg[ c_\theta (7m_\pi^2-4m_K^2) -2\sqrt{2}s_\theta(m_K^2+2m_\pi^2)  \bigg] \,,
\end{eqnarray}
\begin{eqnarray}\label{eq.fetapgg}
F_{\eta'\gamma\gamma} =&& \frac{s_\theta +2\sqrt{2}c_\theta}{4\sqrt{3}\pi^2F} (1+x_{33})+\frac{c_\theta -2\sqrt{2}s_\theta}{4\sqrt{3}\pi^2F} (x_{23}+y_{23}) - \frac{64\sqrt{6}}{3F} c_\theta k_3 
\nonumber \\ && 
-  \frac{64\sqrt{3}}{27F}t_1 \bigg[ s_\theta (7m_\pi^2-4m_K^2) +2\sqrt{2}c_\theta(m_K^2+2m_\pi^2)  \bigg] \,,
\end{eqnarray}
\begin{eqnarray}\label{eq.faxgg}
&& F_{a\gamma\gamma} = \frac{(c_\theta -2\sqrt{2}s_\theta)v_{24}+ (s_\theta +2\sqrt{2}c_\theta)v_{34}}{4\sqrt{3}\pi^2 F} 
\nonumber \\ && 
+\frac{(c_\theta -2\sqrt{2}s_\theta)(x_{24}+y_{24})+(s_\theta +2\sqrt{2}c_\theta)(x_{34}+y_{34})}{4\sqrt{3}\pi^2F}  - \frac{64\sqrt{6}}{3F}k_3 \bigg( \frac{F}{f_a}-\sqrt{6}s_\theta v_{24}+\sqrt{6}c_\theta v_{34} \bigg) 
\nonumber \\ &&
- \frac{64}{27F} t_1 \bigg[-\sqrt{3}s_\theta(4\sqrt{2}v_{24}m_\pi^2-7v_{34}m_\pi^2+  2\sqrt{2}v_{24}m_K^2+4v_{34}m_K^2) 
\nonumber \\ &&
\qquad  \qquad +\sqrt{3}c_\theta(4\sqrt{2}v_{34}m_\pi^2+7v_{24}m_\pi^2+  2\sqrt{2}v_{34}m_K^2-4v_{24}m_K^2) \bigg] \,.
\end{eqnarray}

We use the decay widths of $\pi^0\to\gamma\gamma$, $\eta\to\gamma\gamma$ and $\eta'\to\gamma\gamma$ from the most recent PDG average~\cite{ParticleDataGroup:2022pth} to estimate their two-photon couplings 
\begin{eqnarray}
F_{\pi^0\gamma\gamma}^{\rm Exp} &=& 0.274\pm 0.002 \text{ GeV}^{-1} \,, \\ 
F_{\eta\gamma\gamma}^{\rm Exp} &=&  0.274\pm 0.006 \text{ GeV}^{-1} \,, \\ 
F_{\eta'\gamma\gamma}^{\rm Exp} &=& 0.344\pm 0.008 \text{ GeV}^{-1} \,.
\end{eqnarray}
 Those couplings can be then exploited to determine the NLO LECs from the WZW Lagrangian~\eqref{eq.lagwzwnlo} entering in Eqs.~\eqref{eq.fpigg}-\eqref{eq.fetapgg}. Their explicit values turn out to be
\begin{eqnarray}\label{eq.numt1k3}
t_1= -(4.4 \pm 2.3) \times 10^{-4} {\rm GeV}^{-2} \,, \qquad  k_3= (1.25 \pm 0.23)\times 10^{-4} \,, 
\end{eqnarray}
leading to 
\begin{eqnarray}
  F_{\pi^0\gamma\gamma}&=& 0.276\pm 0.001 \text{ GeV}^{-1}\,, \\
  F_{\eta\gamma\gamma}&=&0.276 \pm 0.009 \text{ GeV}^{-1}\,,  \\ 
F_{\eta'\gamma\gamma}&=&0.343 \pm 0.012 \text{ GeV}^{-1}\,,
\end{eqnarray}
which perfectly reproduce the experimental inputs.

With the values of the NLO LECs of ${\cal L}_{WZW}^{\rm NLO}$ in Eq.~\eqref{eq.numt1k3} and the fitted parameters in Table~\ref{tab.fitnlo}, we can now give our prediction to the two-photon coupling of the axion $F_{a\gamma\gamma}$ up to NLO in the $\delta$ counting
\begin{eqnarray}\label{eq.numfagg}
F_{a\gamma\gamma} = -\frac{[20.1+(0.5\pm 0.1)]\times 10^{-3}}{f_a} \,,
\end{eqnarray}
where the first entry in the numerator on the right hand side corresponds to the LO contribution and the second one denotes the NLO contribution. The two-photon coupling $F_{a\gamma\gamma}$ is related with the $g_{a\gamma\gamma}$ used in Refs.~\cite{GrillidiCortona:2015jxo,Lu:2020rhp} via 
\begin{eqnarray}\label{eq.numgagg}
g_{a\gamma\gamma} = 4\pi \alpha_{em} F_{a\gamma\gamma}= -\frac{\alpha_{em}}{2\pi f_a} \big( 1.63 \pm 0.01 \big) \,, 
\end{eqnarray} 
where the numbers inside the bracket can be compared to the results of $1.92\pm0.04 $~\cite{GrillidiCortona:2015jxo} and $2.05\pm 0.03$~\cite{Lu:2020rhp} from the $SU(2)$ and $SU(3)$ $\chi$PT analyses up to NLO, respectively. The determination of the magnitude of $g_{a\gamma\gamma}$ from the NLO $U(3)$ calculation looks a bit smaller than those from the NLO studies in the $SU(2)$ and $SU(3)$ cases. It is noted that the chiral loops start to appear at NLO in the $SU(2)$ and $SU(3)$ $\chi$PT in the conventional chiral power counting. While, in the $\delta$-counting scheme, the chiral loops only enter at NNLO in the $U(3)$ $\chi$PT. To consistently take into account all the pieces of the two-photon coupling of the axion at NNLO in the $U(3)$ $\chi$PT, one also needs to extend the axion-meson mixing calculations at the same order and we leave this task to a future work.

\section{Summary and Conclusions}\label{sec.summary}

In this work, $U(3)$ chiral perturbation theory is demonstrated to be able to provide a useful framework to calculate the matrix elements of the mixing between the $\pi^0, \eta, \eta'$ and the axion order by order in the joint expansions of the momenta, light-quark masses and $1/N_C$, i.e. the $\delta$-counting scheme. We have performed the complete calculation of the axion-meson mixing and the two-photon couplings of the axion and light-flavor pseudoscalar mesons up to NLO in the $\delta$-counting rule within the framework of the $U(3)$ chiral perturbation theory. The unknown chiral low-energy constants are fixed through the fits to a large amount of lattice QCD data, consisting of the pion-mass dependences of $\eta$-$\eta'$ mixing data,  kaon mass, and the decay constants of the pion and kaon. Reasonable reproductions of the various lattice data are achieved with the $U(3)$ contributions up to NLO. The mixing matrix elements of the $\pi^0, \eta, \eta'$ and the axion are then further used to calculate their two-photon couplings, together with the NLO Wess-Zumino-Witten Lagrangian, where the parameters in the latter Lagrangian are fixed by the experimental two-photon couplings of the $\pi^0, \eta$ and $\eta'$. 

The determined chiral low-energy constants are then used to predict the QCD-axion masses, the mixing strengths of the axion and the $\pi^0, \eta, \eta'$, and the two-photon coupling of the QCD axion. The NLO contributions in the $\delta$ counting to the various axion quantities relative to the LO ones are found to be small. This work paves the way to systematically calculate the interactions between the axion and the light-flavor pseudoscalar mesons $\pi, K, \eta$ and $\eta'$ order by order in the joint expansions of momenta, light-quark masses and $1/N_C$.

\section*{Acknowledgements}
We would like to thank Luca Di Luzio for interesting discussions, and Gunnar Bali and Jakob Simeth for communications to clarify the lattice data of RQCD. This work is funded in part by the Natural Science Foundation of China (NSFC) under Grants Nos.~12150013, 11975090, 12047503, and by the Science Foundation of Hebei Normal University with contract No.~L2023B09. ZHG appreciates the support of Peng Huan-Wu visiting professorship and the hospitality of Institute of Theoretical Physics at Chinese Academy of Sciences, where part of this work has been done. JAO would like to thank partial support by the MICINN AEI (Spain) Grant No. PID2019–106080GB-C22/AEI/10.13039/501100011033, and by the EU Horizon 2020 research and innovation programme, STRONG-2020 project, under Grant agreement No. 824093. RG is partially funded by the Hebei Province Graduate Innovation Funding Project with contract number CXZZBS2023085.

\section*{Appendix: Next-to-leading order coefficients for bilinear terms}

Substituting the LO mixing relations of Eq.~\eqref{eq.mixlo08} into the NLO Lagrangian~\eqref{eq.lagnlo}, one can calculate all the NLO pieces of the bilinear terms in Eq.~\eqref{eq.lagsenlo} and their explicit expressions read

\begin{equation}
\begin{aligned}
\delta^{\pi\eta}_{k}=&\dfrac{-8L_{5}}{3F^{2}}\bigg\{2v_{12}\big[m^{2}_{K}(2 c^{2}_\theta+2\sqrt{2}c_\theta s_\theta+s^{2}_\theta)-m^{2}_{\pi}(2+2\sqrt{2}c_\theta s_\theta-s^{2}_\theta)\big]+\sqrt{3}\epsilon(\sqrt{2}s_\theta-c_\theta)\\ &-2v_{13}(m^{2}_{K}-m^{2}_{\pi})(\sqrt{2}c^{2}_\theta-c_\theta s_\theta-\sqrt{2}s^{2}_\theta)
\bigg\}+\Lambda_{1}s_\theta(c_\theta v_{13}-s_\theta v_{12}),
\end{aligned}
\end{equation}

\begin{equation}
\begin{aligned}
\delta^{\pi\eta}_{m^{2}}=&-\dfrac{16L_{8}}{3F^{2}}\big[4v_{12}m^{2}_{K}(m^{2}_{K}-m^{2}_{\pi})(2c^{2}_\theta+2\sqrt{2}c_\theta s_\theta+s^{2}_\theta)-4v_{13}m^{2}_{K}
(m^{2}_{K}-m^{2}_{\pi})(\sqrt{2}c^{2}_\theta-c_\theta s_\theta\\&-\sqrt{2}s^{2}_\theta)-2\sqrt{3}m^{2}_{\pi}\epsilon(c_\theta-\sqrt{2}s_\theta)\big]-\dfrac{\Lambda_{2}}{3}
\bigg\{2v_{12}s_\theta\big[2m^{2}_{K}(\sqrt{2}c_\theta+s_\theta)+m^{2}_{\pi}(s_\theta-2\sqrt{2}c_\theta)\big]\\&+2v_{13}\big[m^{2}_{K}(-\sqrt{2}c^{2}_\theta-2c_\theta s_\theta+\sqrt{2}s^{2}_\theta)+m^{2}_{\pi}(\sqrt{2}c^{2}_\theta-c_\theta s_\theta-\sqrt{2}s^{2}_\theta)\big]+\sqrt{6}s_\theta\epsilon\bigg\}\,,
\end{aligned}
\end{equation}

\begin{equation}
\begin{aligned}
\delta^{\pi{\eta}'}_{k}=&\dfrac{8L_{5}}{3F^{2}}\bigg\{2v_{12}(m^{2}_{K}-m^{2}_{\pi})(\sqrt{2}c^{2}_\theta-c_\theta s_\theta-\sqrt{2}s^{2}_\theta)+\sqrt{3}\epsilon(\sqrt{2}c_\theta+s_\theta)+2v_{13}\big[-m^{2}_{K}(c^{2}_\theta\\&-2\sqrt{2}c_\theta s_\theta +2s^{2}_\theta)
+m^{2}_{\pi}(2-2\sqrt{2}c_\theta s_\theta-c^{2}_\theta)\big]\bigg\}+\Lambda_{1}c_\theta(-c_\theta v_{13}+s_\theta v_{12})\,,
\end{aligned}
\end{equation}

\begin{equation}
\begin{aligned}
\delta^{\pi{\eta}'}_{m^{2}}=&\dfrac{16L_{8}}{3F^{2}}\big[4v_{12}m^{2}_{K}(m^{2}_{K}-m^{2}_{\pi})(\sqrt{2}c^{2}_\theta-c_\theta s_\theta-\sqrt{2}s^{2}_\theta)
-4v_{13}m^{2}_{K}(m^{2}_{K}-m^{2}_{\pi})(c^{2}_\theta-2\sqrt{2}c_\theta s_\theta+2s^{2}_\theta)\\ &+2\sqrt{3}m^{2}_{\pi}\epsilon(\sqrt{2}c_\theta+s_\theta)\big]
+\dfrac{\Lambda_{2}}{3}\bigg\{2v_{13}c_\theta\big[2m^{2}_{K}(-c_\theta+\sqrt{2}s_\theta)-m^{2}_{\pi}(c_\theta+2\sqrt{2}s_\theta)\big]
\\ &+2v_{12}\big[m^{2}_{K}(\sqrt{2}c^{2}_\theta+2c_\theta s_\theta-
\sqrt{2}s^{2}_\theta)+m^{2}_{\pi}(-\sqrt{2}c^{2}_\theta+c_\theta s_\theta+\sqrt{2}s^{2}_\theta)\big]+\sqrt{6}c_\theta\epsilon\bigg\}\,,
\end{aligned}
\end{equation}

\begin{equation}
\begin{aligned}
\delta^{a}_k=&\dfrac{8L_{5}}{3F^{2}} \bigg\{v^{2}_{24}\big[2m^{2}_{K}(2c^{2}_\theta+2\sqrt{2}c_\theta s_\theta+s^{2}_\theta)+m^{2}_{\pi}(-c^{2}_\theta-4\sqrt{2}c_\theta s_\theta+s^{2}_\theta)\big]+4v_{24}v_{34}\big[m^{2}_{K}(-\sqrt{2}c^{2}_\theta\\& +c_\theta s_\theta+\sqrt{2}s^{2}_\theta)+m^{2}_{\pi}(\sqrt{2}c^{2}_\theta-c_\theta s_\theta
-\sqrt{2}s^{2}_\theta)\big]+v^{2}_{34}\big[2m^{2}_{K}(c^{2}_\theta-2\sqrt{2}c_\theta s_\theta+2s^{2}_\theta)\\ &+m^{2}_{\pi}(c^{2}_\theta+4\sqrt{2}c_\theta s_\theta-s^{2}_\theta)\big]\bigg\}+\dfrac{1}{6}\Lambda_{1}\bigg\{\bigg(\frac{F}{f_a}\bigg)^2+2\sqrt{6}(v_{34}c_\theta-v_{24}s_\theta)\frac{F}{f_a}+6c^{2}_\theta v^{2}_{34}+6s^{2}_\theta v^{2}_{24}\\&
-12c_\theta s_\theta v_{24}v_{34}\bigg\}\,,
\end{aligned}
\end{equation}

\begin{equation}
\begin{aligned}
\delta^{a\pi}_{k}=&\dfrac{8L_{5}}{3F^{2}}\bigg\{-4c_{\theta}^{2}m_{K}^{2}v_{12}v_{24}+m_{\pi}^{2}\big[3v_{14}+s_{\theta}^{2}(-v_{12}v_{24}+2\sqrt{2}v_{13}v_{24}+2\sqrt{2}v_{12}v_{34}+v_{13}v_{34})\\ &+2c_{\theta}s_{\theta}(2\sqrt{2}v_{12}v_{24}+v_{13}v_{24}+v_{12}v_{34}-2\sqrt{2}v_{13}v_{34})+c_{\theta}^{2}(v_{12}v_{24}-2\sqrt{2}v_{13}v_{24}\\ &-2\sqrt{2}v_{12}v_{34}-v_{13}v_{34})\big]+2m_{K}^{2}\big[c_{\theta}^{2}(\sqrt{2}v_{13}v_{24}+\sqrt{2}v_{12}v_{34}-v_{13}v_{34})-s_{\theta}^{2}(v_{12}v_{24}\\ &+\sqrt{2}v_{13}v_{24}+\sqrt{2}v_{12}v_{34}+2v_{13}v_{34})-c_{\theta}s_{\theta}(v_{12}2\sqrt{2}v_{24}+v_{12}v_{34}+v_{13}v_{24}-v_{13}2\sqrt{2}v_{34})\big]\\ &+\sqrt{3}\big[s_{\theta}(-\sqrt{2}v_{24}+v_{34})+c_{\theta}(v_{24}+\sqrt{2}v_{34})\big]\epsilon\bigg\}
+\Lambda_{1}(v_{12}s_\theta-v_{13}c_\theta)\bigg(\dfrac{1}{\sqrt{6}}\dfrac{F}{f_a}-v_{24}s_\theta+v_{34}c_\theta \bigg)\,,
\end{aligned}
\end{equation}

\begin{equation}
\begin{aligned}
\delta^{a\eta}_{k}=&\dfrac{8L_{5}}{3F^{2}}\big[2v_{24}m^{2}_{K}(2c^{2}_\theta+2\sqrt{2}c_\theta s_\theta+s^{2}_\theta)+v_{24}m^{2}_{\pi}(-c^{2}_\theta-4\sqrt{2}c_\theta s_\theta+s^{2}_\theta)-2v_{34}(m^{2}_{K}-m^{2}_{\pi})\\ &(\sqrt{2}c^{2}_\theta-c_\theta s_\theta-\sqrt{2}s^{2}_\theta)\big]+\Lambda_{1}s_\theta \bigg(-\dfrac{1}{\sqrt{6}}\frac{F}{f_a}+v_{24}s_\theta-v_{34}c_\theta \bigg)\,,
\end{aligned}
\end{equation}

\begin{equation}
\begin{aligned}
\delta^{a{\eta}'}_{k}=&\dfrac{-8L_{5}}{3F^{2}} \big[2v_{24}(m^{2}_{K}-m^{2}_{\pi})(\sqrt{2}c^{2}_\theta-c_\theta s_\theta-\sqrt{2}s^{2}_\theta)-
2v_{34}m^{2}_{K}(c^{2}_\theta-2\sqrt{2}c_\theta s_\theta+2s^{2}_\theta)\\ &+v_{34}m^{2}_{\pi}(-c^{2}_\theta-4\sqrt{2}c_\theta s_\theta+s^{2}_\theta)\big]
+\Lambda_{1}c_\theta \bigg(\dfrac{1}{\sqrt{6}}\frac{F}{f_a}-v_{24}s_\theta+v_{34}c_\theta \bigg)\,,
\end{aligned}
\end{equation}

\begin{equation}
\begin{aligned}
\delta_{m^{2}_{a}}=&\frac{16 L_8 }{3 F^2} \bigg\{8 (\sqrt{2} v_{24}^2+v_{34} v_{24}-\sqrt{2}v_{34}^2) c_{\theta } (m_K^2-m_{\pi }^2)
m_K^2 s_{\theta }+c_{\theta }^2\big[4 (2v_{24}^2-2 \sqrt{2} v_{34}v_{24}\\ &+v_{34}^2) m_K^4-4 m_{\pi }^2(2
v_{24}^2-2 \sqrt{2} v_{34} v_{24}+v_{34}^2) m_K^2+3 m_{\pi }^4 (v_{24}^2+v_{34}^2)\big]+s_{\theta
}^2\big[4 (v_{24}^2\\ &+2 \sqrt{2} v_{34} v_{24}+2v_{34}^2) m_K^4-4 m_{\pi }^2(v_{24}^2+2 \sqrt{2} v_{34}
v_{24}+2 v_{34}^2) m_K^2+3 m_{\pi }^4 (v_{24}^2+v_{34}^2)\big]\bigg\}\\  &+\dfrac{\Lambda_{2}}{9}\bigg\{
v_{24}\big[-2\sqrt{6}m^{2}_{K}(\sqrt{2}c_\theta+s_\theta)+\sqrt{6}m^{2}_{\pi}(2\sqrt{2}c_\theta-s_\theta)\big]\frac{F}{f_a}+v_{34}\big[2\sqrt{6}m^{2}_{K}(c_\theta-\sqrt{2}s_\theta)\\ &
+\sqrt{6}m^{2}_{\pi}(c_\theta+2\sqrt{2}s_\theta)\big]\frac{F}{f_a}+6v^{2}_{24}\big[2m^{2}_{K}s_\theta(\sqrt{2}c_\theta+s_\theta)+m^{2}_{\pi}s_\theta(-2\sqrt{2}c_\theta+s_\theta)\big]\\ &
+6v^{2}_{34}\big[2m^{2}_{K}c_\theta(c_\theta-\sqrt{2}s_\theta)+m^{2}_{\pi}c_\theta(c_\theta+2\sqrt{2}s_\theta)\big]
+12v_{24}v_{34}\big[-m^{2}_{K}(\sqrt{2}c^{2}_\theta+2c_\theta s_\theta\\ &-\sqrt{2}s^{2}_\theta)+m^{2}_{\pi}(\sqrt{2}c^{2}_\theta-c_\theta s_\theta-\sqrt{2}s^{2}_\theta)\big]\bigg\}\,,
\end{aligned}
\end{equation}

\begin{equation}
\begin{aligned}
\delta^{a\eta}_{m^{2}}=&\dfrac{-16L_{8}}{3F^{2}}\bigg\{-v_{24}\big[4m^{2}_{K}(m^{2}_{K}-m^{2}_{\pi})(2c^{2}_\theta+2\sqrt{2}c_\theta s_\theta+s^{2}_\theta)+
3m^{4}_{\pi}\big]+4v_{34}m^{2}_{K}(m^{2}_{K}-m^{2}_{\pi})(\sqrt{2}c^{2}_\theta\\ &-c_\theta s_\theta-\sqrt{2}s^{2}_\theta)\bigg\}
-\dfrac{\Lambda_{2}}{18}\bigg\{ \big[ 2\sqrt{6}m^{2}_{K}(\sqrt{2}c_\theta+s_\theta)+\sqrt{6}m^{2}_{\pi}(-2\sqrt{2}c_\theta+s_\theta) \big] \frac{F}{f_a}
\\ & +12v_{24}s_\theta\big[ -2m^{2}_{K}(\sqrt{2}c_\theta+s_\theta)+m^{2}_{\pi}(2\sqrt{2}c_\theta-s_\theta)\big]+
12v_{34}\big[m^{2}_{K}(\sqrt{2}c^{2}_\theta+2c_\theta s_\theta-\sqrt{2}s^{2}_\theta)\\ &+m^{2}_{\pi}(-\sqrt{2}c^{2}_\theta+c_\theta s_\theta+\sqrt{2}s^{2}_\theta)\big]\bigg\}\,,
\end{aligned}
\end{equation}

\begin{equation}
\begin{aligned}
\delta^{a{\eta}'}_{m^{2}}=&\dfrac{-16L_{8}}{3F^{2}}\bigg\{4v_{24}m^{2}_{K}(m^{2}_{K}-m^{2}_{\pi})(\sqrt{2}c^{2}_\theta-c_\theta s_\theta-\sqrt{2}s^{2}_\theta)-v_{34}
\big[4m^{2}_{K}(m^{2}_{K}-m^{2}_{\pi})(c^{2}_\theta-2\sqrt{2}c_\theta s_\theta\\ &+2s^{2}_\theta)+3m^{4}_{\pi}\big]\bigg\}-\dfrac{\Lambda_{2}}{18}\bigg\{ \big[ 2\sqrt{6}m^{2}_{K}(-c_\theta+\sqrt{2}s_\theta)-\sqrt{6}m^{2}_{\pi}(c_\theta+2\sqrt{2}s_\theta) \big] \frac{F}{f_a}
\\ & +12v_{24}\big[m^{2}_{K}(\sqrt{2}c^{2}_\theta +2c_\theta s_\theta-\sqrt{2}s^{2}_\theta)+m^{2}_{\pi}(-\sqrt{2}c^{2}_\theta+c_\theta s_\theta+\sqrt{2}s^{2}_\theta)\big]
\\ & -12v_{34}c_\theta\big[2m^{2}_{K}(c_\theta-\sqrt{2}s_\theta)+m^{2}_{\pi}(c_\theta+2\sqrt{2}s_\theta)\big]\bigg\}\,,
\end{aligned}
\end{equation}

\begin{equation}
\begin{aligned}
\delta^{a\pi}_{m^{2}}=&\frac{16L_8}{3 F^2} \bigg\{-4 m_K^4\big[(2 v_{12} v_{24}-\sqrt{2} v_{13} v_{24}-\sqrt{2} v_{12} v_{34}+v_{13} v_{34}) c_{\theta
}^2+s_{\theta } (v_{12}(2 \sqrt{2} v_{24}+v_{34})\\&+v_{13} (v_{24}-2 \sqrt{2} v_{34})) c_{\theta
}+s_{\theta }^2 (v_{12} v_{24}+\sqrt{2} v_{13} v_{24}+\sqrt{2} v_{12} v_{34}+2 v_{13} v_{34})\big]\\&+2 m_{\pi }^2\big[2
c_{\theta }^2 (2 v_{12} v_{24}-\sqrt{2} v_{13} v_{24}-\sqrt{2} v_{12} v_{34}+v_{13} v_{34}) m_K^2\\&+2 c_{\theta } s_{\theta }
(2 \sqrt{2} v_{12} v_{24}+v_{13} v_{24}+v_{12} v_{34}-2 \sqrt{2} v_{13} v_{34}) m_K^2+\sqrt{3} \epsilon  c_{\theta }
(v_{24}+\sqrt{2} v_{34})\\&+s_{\theta } (2 s_{\theta } (v_{12} v_{24}+\sqrt{2} v_{13} v_{24}+\sqrt{2} v_{12}
v_{34}+2 v_{13} v_{34}) m_K^2+\sqrt{3} \epsilon  (v_{34}-\sqrt{2} v_{24}))\big]\\&+3 m_{\pi
}^4\big[v_{14}-(c_{\theta }^2+s_{\theta }^2) (v_{12} v_{24}+v_{13} v_{34})\big]\bigg\}-\dfrac{\Lambda_{2}}{18}\bigg\{-6\epsilon\bigg(\frac{F}{f_a}-\sqrt{6}s_\theta v_{24}+\sqrt{6}c_\theta v_{34} \bigg)\\ &
+v_{12}\big[-2\sqrt{6}m^{2}_{K}(\sqrt{2}c_\theta+s_\theta)+\sqrt{6}m^{2}_{\pi}(2\sqrt{2}c_\theta-s_\theta)\big]\frac{F}{f_a}+v_{13}\big[2\sqrt{6}m^{2}_{K}(c_\theta-\sqrt{2}s_\theta)\\ & +\sqrt{6}m^{2}_{\pi}(c_\theta+2\sqrt{2}s_\theta)\big]\frac{F}{f_a}+12v_{12}v_{24}s_\theta\big[2m^{2}_{K}(\sqrt{2}c_\theta+s_\theta)+m^{2}_{\pi}(-2\sqrt{2}c_\theta+s_\theta)\big]\\ & +
12(v_{12}v_{34}+v_{13}v_{24})\big[-m^{2}_{K}(\sqrt{2}c^{2}_\theta+2c_\theta s_\theta-\sqrt{2}s^{2}_\theta)+m^{2}_{\pi}(\sqrt{2}c^{2}_\theta-c_\theta s_\theta-\sqrt{2}s^{2}_\theta)\big]\\ &+12v_{13}v_{34}c_\theta\big[2m^{2}_{K}(c_\theta-\sqrt{2}s_\theta)+m^{2}_{\pi}(c_\theta+2\sqrt{2}s_\theta)\big]\bigg\}\,.
\end{aligned}
\end{equation}

\begin{equation}\label{eq.deltakpi}
\begin{aligned}
\delta^{\pi}_k=\dfrac{8L_{5}m^{2}_{\pi}}{F^{2}}\,,
\end{aligned}
\end{equation}

\begin{equation}
\begin{aligned}
\delta_{m^{2}_{\pi}}=\dfrac{16L_{8}m^{4}_{\pi}}{F^{2}}\,,
\end{aligned}
\end{equation}

\begin{equation} 
\begin{aligned}
\delta^{K}_k=\dfrac{8L_{5}m^{2}_{K}}{F^{2}}\,,
\end{aligned}
\end{equation}

\begin{equation}
\begin{aligned}
\delta_{m^{2}_{K}}=\dfrac{16L_{8}m^{4}_{K}}{F^{2}}\,,
\end{aligned}
\end{equation}

\begin{equation}
 \delta^{\eta}_k=\dfrac{8 L_{5}[c_{\theta}^{2}(4m_{K}^{2}-m_{\pi}^{2})+
4\sqrt{2}c_{\theta}(m_{K}^{2}-m_{\pi}^{2})s_{\theta}+(2m_{K}^{2}+m_{\pi}^{2})s_{\theta}^{2}]}{3F^{2}} +s_{\theta}^{2}\Lambda_{1}\,,
\end{equation}

\begin{equation}
\delta^{\eta'}_k= \dfrac{8 L_{5}[c_{\theta}^{2}(2m_{K}^{2}
+m_{\pi}^{2})+4\sqrt{2}c_{\theta}(-m_{K}^{2}+m_{\pi}^{2})s_{\theta}+(4m_{K}^{2}-m_{\pi}^{2})s_{\theta}^{2}]}{3F^{2}}+ c_{\theta}^{2}\Lambda_{1}\,,
\end{equation}

\begin{equation} 
\delta^{\eta\eta'}_{k}=  -\dfrac{16 L_{5}(m_{K}^{2}-m_{\pi}^{2})
(\sqrt{2}c_{\theta}^{2}-c_{\theta}s_{\theta}-\sqrt{2}s_{\theta}^{2})}{3F^{2}}-c_{\theta}s_{\theta}\Lambda_{1}\,,
\end{equation}

\begin{equation}
\begin{aligned}
 \delta_{m_{\eta}^{2}}=& \dfrac{16 L_{8}}{3F^{2}}[c_{\theta}^{2}(8m_{K}^{4}-8m_{K}^{2}m_{\pi}^{2}+3m_{\pi}^{4})+8\sqrt{2}c_{\theta}m_{K}^{2}
(m_{K}^{2}-m_{\pi}^{2})s_{\theta}+(4m_{K}^{4}-4m_{K}^{2}m_{\pi}^{2}+3m_{\pi}^{4})s_{\theta}^{2}]  \\ &
 +\dfrac{2}{3}s_{\theta}[2\sqrt{2}c_{\theta}(m_{K}^{2}-m_{\pi}^{2})+(2m_{K}^{2}+m_{\pi}^{2})s_{\theta}]\Lambda_{2} \,,
\end{aligned}
\end{equation}

\begin{equation}
\begin{aligned}
\delta_{m_{\eta'}^{2}}=&  \dfrac{16 L_{8}}{3F^{2}}[c_{\theta}^{2}(4m_{K}^{4}-4m_{K}^{2}m_{\pi}^{2}+3m_{\pi}^{4})+8\sqrt{2}c_{\theta}
m_{K}^{2}(-m_{K}^{2}+m_{\pi}^{2})s_{\theta} +(8m_{K}^{4}-8m_{K}^{2}m_{\pi}^{2}+3m_{\pi}^{4})s_{\theta}^{2}] \\ &
+ \dfrac{2}{3}c_{\theta}[c_{\theta}(2m_{K}^{2}+
m_{\pi}^{2})+2\sqrt{2}(-m_{K}^{2}+m_{\pi}^{2})s_{\theta}]\Lambda_{2} \,,
\end{aligned}
\end{equation}

\begin{equation}\label{eq.deltam2etaetap}
\begin{aligned}
 \delta^{\eta\eta'}_{m^{2}}=& -\dfrac{64 L_{8}m_{K}^{2}(m_{K}^{2}-m_{\pi}^{2})(\sqrt{2}c_{\theta}^{2}-c_{\theta}s_{\theta}-\sqrt{2}
s_{\theta}^{2})}{3F^{2}} \\&
-\dfrac{2}{3}[\sqrt{2}c_{\theta}^{2}(m_{K}^{2}-m_{\pi}^{2})+c_{\theta}
(2m_{K}^{2}+m_{\pi}^{2})s_{\theta}+\sqrt{2}(-m_{K}^{2}+m_{\pi}^{2})s_{\theta}^{2}]\Lambda_{2}\,.
\end{aligned}
\end{equation}
Although the NLO formulas in Eqs.~\eqref{eq.deltakpi}-\eqref{eq.deltam2etaetap} can be found in Ref.~\cite{Guo:2015xva}, we show their explicit expressions here for the sake of completeness.


\begin{thebibliography}{90}


\bibitem{Peccei:1977hh}
R.~D.~Peccei and H.~R.~Quinn,
Phys. Rev. Lett. \textbf{38}, 1440-1443 (1977)
doi:10.1103/PhysRevLett.38.1440

\bibitem{Weinberg:1977ma}
S.~Weinberg,
Phys. Rev. Lett. \textbf{40}, 223-226 (1978)
doi:10.1103/PhysRevLett.40.223

\bibitem{Wilczek:1977pj}
F.~Wilczek,
Phys. Rev. Lett. \textbf{40}, 279-282 (1978)
doi:10.1103/PhysRevLett.40.279

\bibitem{axionreviews}
J.~E.~Kim and G.~Carosi,
Rev. Mod. Phys. \textbf{82}, 557-602 (2010)
[erratum: Rev. Mod. Phys. \textbf{91}, no.4, 049902 (2019)]
doi:10.1103/RevModPhys.82.557
[arXiv:0807.3125 [hep-ph]].

L.~Di Luzio, M.~Giannotti, E.~Nardi and L.~Visinelli,
Phys. Rept. \textbf{870}, 1-117 (2020)
doi:10.1016/j.physrep.2020.06.002
[arXiv:2003.01100 [hep-ph]].


P.~W.~Graham, I.~G.~Irastorza, S.~K.~Lamoreaux, A.~Lindner and K.~A.~van Bibber,
Ann. Rev. Nucl. Part. Sci. \textbf{65}, 485-514 (2015)
doi:10.1146/annurev-nucl-102014-022120
[arXiv:1602.00039 [hep-ex]].

I.~G.~Irastorza and J.~Redondo,
Prog. Part. Nucl. Phys. \textbf{102}, 89-159 (2018)
doi:10.1016/j.ppnp.2018.05.003
[arXiv:1801.08127 [hep-ph]].

P.~Sikivie,
Rev. Mod. Phys. \textbf{93}, no.1, 015004 (2021)
doi:10.1103/RevModPhys.93.015004
[arXiv:2003.02206 [hep-ph]].

K.~Choi, S.~H.~Im and C.~Sub Shin,
Ann. Rev. Nucl. Part. Sci. \textbf{71}, 225-252 (2021)
doi:10.1146/annurev-nucl-120720-031147
[arXiv:2012.05029 [hep-ph]].

\bibitem{Dine:1981rt}
M.~Dine, W.~Fischler and M.~Srednicki,
Phys. Lett. B \textbf{104}, 199-202 (1981)
doi:10.1016/0370-2693(81)90590-6

\bibitem{Zhitnitsky:1980tq}
A.~R.~Zhitnitsky,
Sov. J. Nucl. Phys. \textbf{31}, 260 (1980)

\bibitem{Kim:1979if}
J.~E.~Kim,
Phys. Rev. Lett. \textbf{43}, 103 (1979)
doi:10.1103/PhysRevLett.43.103

\bibitem{Shifman:1979if}
M.~A.~Shifman, A.~I.~Vainshtein and V.~I.~Zakharov,
Nucl. Phys. B \textbf{166}, 493-506 (1980)
doi:10.1016/0550-3213(80)90209-6




\bibitem{Choi:1986zw}
K.~Choi, K.~Kang and J.~E.~Kim,
Phys. Lett. B \textbf{181}, 145-149 (1986)
doi:10.1016/0370-2693(86)91273-6

\bibitem{Spalinski:1988yf}
M.~Spalinski,
J. Phys. G \textbf{14}, L67-L70 (1988)
doi:10.1088/0305-4616/14/5/001

\bibitem{Alves:2020xhf}
D.~S.~M.~Alves,
Phys. Rev. D \textbf{103}, no.5, 055018 (2021)
doi:10.1103/PhysRevD.103.055018
[arXiv:2009.05578 [hep-ph]].

\bibitem{Bigazzi:2019hav}
F.~Bigazzi, A.~L.~Cotrone, M.~J\"arvinen and E.~Kiritsis,
JHEP \textbf{01}, 100 (2020)
doi:10.1007/JHEP01(2020)100
[arXiv:1906.12132 [hep-ph]].

\bibitem{DiVecchia:2017xpu}
P.~Di Vecchia, G.~Rossi, G.~Veneziano and S.~Yankielowicz,
JHEP \textbf{12}, 104 (2017)
doi:10.1007/JHEP12(2017)104
[arXiv:1709.00731 [hep-th]].


\bibitem{Ertas:2020xcc}
F.~Ertas and F.~Kahlhoefer,
JHEP \textbf{07}, 050 (2020)
doi:10.1007/JHEP07(2020)050
[arXiv:2004.01193 [hep-ph]].

\bibitem{Gan:2020aco}
L.~Gan, B.~Kubis, E.~Passemar and S.~Tulin,
Phys. Rept. \textbf{945}, 1-105 (2022)
doi:10.1016/j.physrep.2021.11.001
[arXiv:2007.00664 [hep-ph]].

\bibitem{Landini:2019eck}
G.~Landini and E.~Meggiolaro,
Eur. Phys. J. C \textbf{80}, no.4, 302 (2020)
doi:10.1140/epjc/s10052-020-7849-2
[arXiv:1906.03104 [hep-ph]].

\bibitem{Bottaro:2020dqh}
S.~Bottaro and E.~Meggiolaro,
Phys. Rev. D \textbf{102}, no.1, 014048 (2020)
doi:10.1103/PhysRevD.102.014048
[arXiv:2004.11901 [hep-ph]].


\bibitem{REDTOP:2022slw}
J.~Elam \textit{et al.} [REDTOP],
[arXiv:2203.07651 [hep-ex]].


\bibitem{Aloni:2018vki}
D.~Aloni, Y.~Soreq and M.~Williams,
Phys. Rev. Lett. \textbf{123}, no.3, 031803 (2019)
doi:10.1103/PhysRevLett.123.031803
[arXiv:1811.03474 [hep-ph]].

\bibitem{Alves:2017avw}
D.~S.~M.~Alves and N.~Weiner,
JHEP \textbf{07}, 092 (2018)
doi:10.1007/JHEP07(2018)092
[arXiv:1710.03764 [hep-ph]].

\bibitem{Herrera-Siklody:1996tqr}
P.~Herrera-Siklody, J.~I.~Latorre, P.~Pascual and J.~Taron,
Nucl. Phys. B \textbf{497}, 345-386 (1997)
doi:10.1016/S0550-3213(97)00260-5
[arXiv:hep-ph/9610549 [hep-ph]].


\bibitem{Leutwyler:1997yr}
H.~Leutwyler,
Nucl. Phys. B Proc. Suppl. \textbf{64}, 223-231 (1998)
doi:10.1016/S0920-5632(97)01065-7
[arXiv:hep-ph/9709408 [hep-ph]].

\bibitem{Kaiser:2000gs}
R.~Kaiser and H.~Leutwyler,
Eur. Phys. J. C \textbf{17}, 623-649 (2000)
doi:10.1007/s100520000499
[arXiv:hep-ph/0007101 [hep-ph]].

\bibitem{Dimopoulos:1979pp}
S.~Dimopoulos,
Phys. Lett. B \textbf{84}, 435-439 (1979)
doi:10.1016/0370-2693(79)91233-4

\bibitem{Holdom:1982ex}
B.~Holdom and M.~E.~Peskin,
Nucl. Phys. B \textbf{208}, 397-412 (1982)
doi:10.1016/0550-3213(82)90228-0

\bibitem{Rubakov:1997vp}
V.~A.~Rubakov,
JETP Lett. \textbf{65}, 621-624 (1997)
doi:10.1134/1.567390
[arXiv:hep-ph/9703409 [hep-ph]].

\bibitem{Gaillard:2018xgk}
M.~K.~Gaillard, M.~B.~Gavela, R.~Houtz, P.~Quilez and R.~Del Rey,
Eur. Phys. J. C \textbf{78}, no.11, 972 (2018)
doi:10.1140/epjc/s10052-018-6396-6
[arXiv:1805.06465 [hep-ph]].


\bibitem{Georgi:1986df}
H.~Georgi, D.~B.~Kaplan and L.~Randall,
Phys. Lett. B \textbf{169}, 73-78 (1986)
doi:10.1016/0370-2693(86)90688-X

\bibitem{GrillidiCortona:2015jxo}
G.~Grilli di Cortona, E.~Hardy, J.~Pardo Vega and G.~Villadoro,
JHEP \textbf{01}, 034 (2016)
doi:10.1007/JHEP01(2016)034
[arXiv:1511.02867 [hep-ph]].
%

\bibitem{ua1nc} E.~Witten, Nucl.\ Phys.\ B {\bf 156}, 269 (1979); 
S.~Coleman and E.~Witten, Phys.~Rev.~Lett.~{\bf 45}, 100 (1980);
G.~Veneziano, Nucl.\ Phys.\ B {\bf 159}, 213 (1979).


\bibitem{Gasser:1984gg}
J.~Gasser and H.~Leutwyler,
Nucl. Phys. B \textbf{250}, 465-516 (1985)
doi:10.1016/0550-3213(85)90492-4

\bibitem{coleman69} S. R. Coleman, J. Wess and B. Zumino, Phys. Rev. {\bf 177}, 2239-2247 (1969). 

\bibitem{Wess:1971yu}
J.~Wess and B.~Zumino,
Phys. Lett. B \textbf{37}, 95-97 (1971)
doi:10.1016/0370-2693(71)90582-X

\bibitem{Witten:1983tw}
E.~Witten,
Nucl. Phys. B \textbf{223}, 422-432 (1983)
doi:10.1016/0550-3213(83)90063-9

\bibitem{Moussallam:1994xp}
B.~Moussallam,
Phys. Rev. D \textbf{51}, 4939-4949 (1995)
doi:10.1103/PhysRevD.51.4939
[arXiv:hep-ph/9407402 [hep-ph]].

\bibitem{Bijnens:2001bb}
J.~Bijnens, L.~Girlanda and P.~Talavera,
Eur. Phys. J. C \textbf{23}, 539-544 (2002)
doi:10.1007/s100520100887
[arXiv:hep-ph/0110400 [hep-ph]].
 
\bibitem{Jamin:2000wn}
  M.~Jamin, J.A. Oller and A. Pich, 
  Nucl. Phys. B \textbf{587}, 331--362 (2000)
  doi:10.1016/S0550-3213(00)00479-X
  [arXiv:hep-ph/0006045 [hep-ph]].

\bibitem{Jamin:2001zq}
  M.~Jamin, J.A. Oller and A. Pich, 
  Nucl. Phys. B \textbf{622}, 279--308 (2002)
  doi:10.1016/S0550-3213(01)00605-8
  [arXiv:hep-ph/0110193 [hep-ph]].


\bibitem{Guo:2011pa}
Z.~H.~Guo and J.~A.~Oller,
Phys. Rev. D \textbf{84}, 034005 (2011)
doi:10.1103/PhysRevD.84.034005
[arXiv:1104.2849 [hep-ph]].

\bibitem{Guo:2012ym}
Z.~H.~Guo, J.~A.~Oller and J.~Ruiz de Elvira,
Phys. Lett. B \textbf{712}, 407-412 (2012)
doi:10.1016/j.physletb.2012.05.021
[arXiv:1203.4381 [hep-ph]].

\bibitem{Guo:2012yt}
Z.~H.~Guo, J.~A.~Oller and J.~Ruiz de Elvira,
Phys. Rev. D \textbf{86}, 054006 (2012)
doi:10.1103/PhysRevD.86.054006
[arXiv:1206.4163 [hep-ph]].




\bibitem{Guo:2015xva}
X.~K.~Guo, Z.~H.~Guo, J.~A.~Oller and J.~J.~Sanz-Cillero,
JHEP \textbf{06}, 175 (2015)
doi:10.1007/JHEP06(2015)175
[arXiv:1503.02248 [hep-ph]].



\bibitem{Dashen:1969eg}
R.~F.~Dashen,
Phys. Rev. \textbf{183}, 1245-1260 (1969)
doi:10.1103/PhysRev.183.1245

\bibitem{Gu:2018swy}
X.~W.~Gu, C.~G.~Duan and Z.~H.~Guo,
Phys. Rev. D \textbf{98}, no.3, 034007 (2018)
doi:10.1103/PhysRevD.98.034007
[arXiv:1803.07284 [hep-ph]].
 
\bibitem{Lu:2020rhp}
Z.~Y.~Lu, M.~L.~Du, F.~K.~Guo, U.~G.~Mei\ss{}ner and T.~Vonk,
JHEP \textbf{05}, 001 (2020)
doi:10.1007/JHEP05(2020)001
[arXiv:2003.01625 [hep-ph]].


\bibitem{Ottnad:2017bjt}
K.~Ottnad \textit{et al.} [ETM],
Phys. Rev. D \textbf{97}, no.5, 054508 (2018)
doi:10.1103/PhysRevD.97.054508
[arXiv:1710.07986 [hep-lat]].


\bibitem{Gregory:2011sg}
E.~B.~Gregory \textit{et al.} [UKQCD],
Phys. Rev. D \textbf{86}, 014504 (2012)
doi:10.1103/PhysRevD.86.014504
[arXiv:1112.4384 [hep-lat]].

\bibitem{Christ:2010dd}
N.~H.~Christ, C.~Dawson, T.~Izubuchi, C.~Jung, Q.~Liu, R.~D.~Mawhinney, C.~T.~Sachrajda, A.~Soni and R.~Zhou,
Phys. Rev. Lett. \textbf{105}, 241601 (2010)
doi:10.1103/PhysRevLett.105.241601
[arXiv:1002.2999 [hep-lat]].

\bibitem{Dudek:2011tt}
J.~J.~Dudek, R.~G.~Edwards, B.~Joo, M.~J.~Peardon, D.~G.~Richards and C.~E.~Thomas,
Phys. Rev. D \textbf{83}, 111502 (2011)
doi:10.1103/PhysRevD.83.111502
[arXiv:1102.4299 [hep-lat]].

\bibitem{Bali:2021qem}
G.~S.~Bali \textit{et al.} [RQCD],
JHEP \textbf{08}, 137 (2021)
doi:10.1007/JHEP08(2021)137
[arXiv:2106.05398 [hep-lat]].


\bibitem{Chen:2012vw}
Y.~H.~Chen, Z.~H.~Guo and H.~Q.~Zheng,
Phys. Rev. D \textbf{85}, 054018 (2012)
doi:10.1103/PhysRevD.85.054018
[arXiv:1201.2135 [hep-ph]].

\bibitem{Chen:2014yta}
Y.~H.~Chen, Z.~H.~Guo and B.~S.~Zou,
Phys. Rev. D \textbf{91}, 014010 (2015)
doi:10.1103/PhysRevD.91.014010
[arXiv:1411.1159 [hep-ph]].

\bibitem{RBC:2010qam}
Y.~Aoki \textit{et al.} [RBC and UKQCD],
Phys. Rev. D \textbf{83}, 074508 (2011)
doi:10.1103/PhysRevD.83.074508
[arXiv:1011.0892 [hep-lat]].

\bibitem{RBC:2012cbl}
R.~Arthur \textit{et al.} [RBC and UKQCD],
Phys. Rev. D \textbf{87}, 094514 (2013)
doi:10.1103/PhysRevD.87.094514
[arXiv:1208.4412 [hep-lat]].


\bibitem{Durr:2010hr}
S.~Durr, Z.~Fodor, C.~Hoelbling, S.~D.~Katz, S.~Krieg, T.~Kurth, L.~Lellouch, T.~Lippert, A.~Ramos and K.~K.~Szabo,
Phys. Rev. D \textbf{81}, 054507 (2010)
doi:10.1103/PhysRevD.81.054507
[arXiv:1001.4692 [hep-lat]].

 
\bibitem{ParticleDataGroup:2022pth}
R.~L.~Workman \textit{et al.} [Particle Data Group],
PTEP \textbf{2022}, 083C01 (2022)
doi:10.1093/ptep/ptac097
 
  
\end{thebibliography}
\end{document}